\documentclass[twocolumn]{aastex61}
\usepackage{natbib}
\defcitealias{Rackham2018}{Paper I}
\usepackage{outlines}
\usepackage{enumitem}
\setenumerate[1]{label=\arabic*.}
\setenumerate[2]{label=\Alph*)}
\setenumerate[3]{label=\roman*)}
\setenumerate[4]{label=\alph*)}
\setenumerate[5]{label=\arabic*)}

\newcommand{\ponet}{\citetalias{Rackham2018}}
\newcommand{\ponep}{\citepalias{Rackham2018}}
\newcommand{\spots}{\texttt{spots}}
\newcommand{\spotsfaculae}{\texttt{spots+faculae}}

\received{2018 September 7}
\revised{2018 December 12}
\accepted{2018 December 14}
\submitjournal{The Astronomical Journal}

\shorttitle{Transit Light Source Effect II}
\shortauthors{Rackham, Apai, \& Giampapa}

\begin{document}

\title{The Transit Light Source Effect II: The Impact of Stellar Heterogeneity on Transmission Spectra of Planets Orbiting Broadly Sun-like Stars}

\correspondingauthor{Benjamin V. Rackham}
\email{brackham@as.arizona.edu}

\author{Benjamin V. Rackham}
\affiliation{Department of Astronomy/Steward Observatory, The University of Arizona, 933 N. Cherry Avenue, Tucson, AZ 85721, USA}
\affiliation{Earths in Other Solar Systems Team, NASA Nexus for Exoplanet System Science.}
\affiliation{National Science Foundation Graduate Research Fellow.}

\author{D\'aniel Apai}
\affiliation{Department of Astronomy/Steward Observatory, The University of Arizona, 933 N. Cherry Avenue, Tucson, AZ 85721, USA}
\affiliation{Department of Planetary Sciences, The University of Arizona, 1629 E. University Blvd, Tucson, AZ 85721, USA}
\affiliation{Max Planck Institute for Astronomy, K\"onigstuhl 17, 69117 Heidelberg, Germany}
\affiliation{Earths in Other Solar Systems Team, NASA Nexus for Exoplanet System Science.}

\author{Mark S. Giampapa}
\affiliation{National Solar Observatory, 950 N. Cherry Avenue, Tucson, AZ 85719, USA}
\affiliation{Lunar and Planetary Laboratory, University of Arizona, Tucson, AZ 85711, USA}



\begin{abstract}
Transmission spectra probe exoplanetary atmospheres, but they can also be strongly affected by heterogeneities in host star photospheres through the transit light source effect.
Here we build upon our recent study of the effects of unocculted spots and faculae on M-dwarf transmission spectra, extending the analysis to FGK dwarfs.
Using a suite of rotating model photospheres, we explore spot and 
\replaced{faculae}
{facula}
covering fractions for varying activity levels and the associated stellar contamination spectra.
Relative to M dwarfs, we find that the typical variabilities of FGK dwarfs imply lower spot covering fractions, though they generally increase with later spectral types, from 
\replaced{roughly 0.1\%}{$\sim$0.1\%}
for F dwarfs to 2--4\% for late-K dwarfs.
While the stellar contamination spectra are considerably weaker than those for typical M dwarfs, we find that typically active G and K dwarfs produce visual slopes that are detectable in high-precision transmission spectra.
We examine line offsets at H$\alpha$ and the Na and K doublets and find that unocculted faculae in K dwarfs can appreciably alter transit depths around the Na D doublet.
We find that band-averaged transit depth offsets at molecular bands for CH$_{4}$, CO, CO$_{2}$, H$_{2}$O, N$_{2}$O, O$_{2}$, and O$_{3}$ are not detectable for typically active FGK dwarfs, though stellar TiO/VO features are potentially detectable for typically active late-K dwarfs.
Generally, this analysis shows that 
\replaced{typical}{inactive}
FGK dwarfs do not produce detectable stellar contamination features in transmission spectra, though 
\added{active FGK host stars can produce such features and}
care is warranted in 
\replaced{dealing with more active FGK host stars.}{interpreting transmission spectra from these systems.}
\end{abstract}




\section{Introduction} \label{sec:intro}

Transiting exoplanets provide an opportunity to study the atmospheres of distant worlds.
During a transit, the host star illuminates the exoplanet's atmosphere, enabling measurements of the properties of the optically thin upper atmosphere.
Changes in transit depth as a function of wavelength, i.e. the transmission spectrum, encode information about absorption and scattering in the exoplanet's atmosphere \citep{Seager2000, Brown2001, Hubbard2001}.
This technique has led to discoveries of atomic and molecular absorption in exoplanetary atmospheres \citep[e.g.,][]{Charbonneau2002, Sing2012}, provided constraints on their bulk metallicities \citep{Fraine2014, Kreidberg2014b, Kreidberg2015, Wakeford2017, Wakeford2018, Nikolov2018}, and has recently begun to enable comparative studies of exoplanetary atmospheres \citep{Sing2016, Barstow2017, Pinhas2018, Pinhas2019}.

At the same time, photospheric heterogeneities \deleted{, i.e. spots and faculae,} on the host star produce wavelength-dependent effects on the transmission spectrum through the transit light source (TLS) effect.
Essentially, transit observations are differential measurements that necessarily compare transit depth changes to an out-of-transit baseline.
However, the out-of-transit baseline is set by the integrated stellar disk, while the actual light source for the transmission measurement is provided by the emergent spectrum of the spatially resolved transit chord.
As a result, any spectral difference between the integrated stellar disk and the transit chord will be imprinted on the differential measurement \citep{Pont2008, Berta2011, Sing2011, McCullough2014}.
Given this fundamental difference from a classical laboratory transmission measurement, in which the spectrum of the light source is well-characterized, exoplanet transmission spectroscopy studies should assume some level of TLS contamination (or ``stellar'' contamination) exists with every measurement and seek to place limits on it.
For further context on stellar contamination of transmission spectra, we refer the reader to \citet{Apai2018}.

\added{The most prominent photospheric heterogeneities are magnetic active regions.
These include spots---cool, dark regions of suppressed convection \citep{Parker1955, Babcock1961}---and faculae---the hot, bright walls of flux tubes \citep{Spruit1976} and granules \citep{Keller2004, Lites2004} revealed via projection effects.
These active regions, i.e. spots and faculae, are ubiquitous features of stars with convective outer layers \citep[See reviews by][]{Ruzmaikin2001, Berdyugina2005, Strassmeier2009, Collier_Cameron2017}.
When present within the transit chord, active regions produce time-resolved bumps in transit light curves that affect transit depth determinations \citep[e.g.,][]{Pont2008}.
More perniciously, when present outside the transit chord, active regions affect transit depths through the TLS effect.}

The ability of stars to imprint spectral features in transmission spectra has been recognized for more than a decade, mostly in the form of in-depth studies of individual exoplanet host stars \citep[e.g., ][]{Pont2008, Pont2013, Bean2010, Berta2011, Sing2011, Jordan2013, Oshagh2014, Cauley2017, Rackham2017}.
In a systematic study of stellar contamination in M dwarfs \citep[][hereafter Paper I]{Rackham2018}, we found that rotational variability amplitudes that are typically observed for M dwarfs correspond to a wide range of spot and 
\replaced{faculae}
{facula}
covering fractions.
Accordingly, a wide uncertainty exists for the scale of the stellar contamination spectra associated with these active regions.
This finding has important implications for high-precision observations of low-mass planets around M dwarfs, for which active regions can imprint molecular features in transmission spectra on a scale that is comparable to or even an order of magnitude larger than that of atmospheric features of small, rocky exoplanets. 

In contrast to their M-dwarf counterparts, FGK dwarfs generally display lower relative amplitudes of rotational brightness variations.
For example, \citet{McQuillan2014} derive rotation periods for over 34,000 main-sequence \textit{Kepler} stars with effective temperatures below 6500~K, roughly a quarter of the full \textit{Kepler} sample.
They find that periodic fractions decreases with increasing temperature, from 0.83 for stars in their coolest temperature bin ($T_\mathrm{eff} < 4000$~K) to 0.20 for stars in their hottest bin ($T_\mathrm{eff} \in [6000, 6500]$~K), in broad agreement with results from \citet{Basri2013}.
They also find that variability amplitudes decrease with increasing temperature as well, with median amplitudes of 0.7\% and 0.2\% for these same bins (see their Table~1 and Figure 3).
These lower periodic fractions and variability amplitudes for hotter stars point to differences in the properties of magnetic active regions and suggest that FGK dwarfs generally pose fewer difficulties for transmission spectroscopy observations than cooler stars.

Yet, despite their overall lower rotational variabilities, FGK stars still present their own challenges for transmission spectroscopy.
Recently, \citet{Cauley2018} examined the effects of spots and faculae on chromospherically sensitive atomic lines in high-resolution visual transmission spectra of G and K dwarfs.
They explored models for four effective temperatures from 4500 K to 6000 K, corresponding to mid-K to early-G dwarfs, and ranges of spot covering fractions from 0.3\% to 10\% and facular covering fractions from 5\ to 50\%.
They found that large facular covering fractions can appreciably alter transit depths for H$\alpha$, Ca II K, and Na I D, which underscores the need to constrain active region covering fractions for active G and K dwarf hosts in order to properly interpret atomic line detections in high-resolution transmission spectra.

Observational efforts also attest to the challenges posed by FGK stars to transmission spectroscopy studies.
The hot Jupiter \object{HD~189733b}, for example, demonstrates a strong blue-ward slope in its visual transmission spectrum, which has been interpreted as Rayleigh scattering by condensate grains in the planetary atmosphere \citep{Lecavelier2008, Pont2008, Pont2013, Sing2011, Sing2016}.
The data used to arrive at this interpretation have been corrected for the effect of the 1--2\% coverage of unocculted spots that one would infer from variability monitoring of the K1V host star \citep{Pont2013}.
However, if the spot coverage is instead $\sim 4\%$, \citet{McCullough2014} showed that the observed transmission spectrum is also consistent with a clear planetary atmosphere and a larger contribution from unocculted spots.
Furthermore, some uncertainty exists as to whether the in-transit H$\alpha$ absorption signature has a stellar or planetary origin or some combination of both, though the lack of a clear relationship between the stellar activity level and the H$\alpha$ absorption signal argues against a purely stellar origin \citep{Cauley2017}.
Similarly, in a recent study of the visual transmission spectrum of the hot Jupiter \object{WASP-19b}, \citet{Espinoza2019} applied an atmospheric retrieval approach that considers both stellar and planetary spectral features and found that the TiO features observed in one of their six transits likely originate with unocculted spots on the active G9V host star, in contrast to previous planetary interpretations for the features \citep{Sedaghati2017}.
These tensions in interpretations illustrate the need for a systematic study of the spectral features produced in transmission spectra by broadly Sun-like stars.

In this work, we extend our analysis of the TLS effect to investigate stellar contamination in 0.05--5.5~$\micron$ transmission spectra of exoplanets with FGK host stars.
We find that stellar contamination is generally less problematic for FGK dwarfs, though potentially observable signals are possible for more active host stars, later spectral types, and observations at shorter wavelengths.
Section~\ref{sec:variability} details the rotational variability model for FGK dwarfs that we use to determine spot and facula covering fractions corresponding to typical activity levels.
We present in Section~\ref{sec:contamination} the contamination spectra for typically active FGK dwarfs. 
In Section~\ref{sec:discussion}, we discuss the scale of the stellar contamination and examine trends in spectral features, and we summarize the key findings of this analysis in Section~\ref{sec:conclusions}.

\section{Stellar Variability Modeling} \label{sec:variability}

We modeled rotational variability amplitudes due to photospheric heterogeneities for F, G, and K dwarfs following the method detailed in \ponet{}. 
\added{Following convention, we organized our analysis around spectral types, which is effectively the same as organizing it by effective temperature with irregular grid spacing.}
In the following section, we briefly summarize the methodology and detail differences in the current analysis with respect to \ponet{}.

\subsection{Adopted Stellar Parameters}

We generated model photospheres for spectral types F5V--K9V, including three photospheric components---immaculate photosphere, spots, and faculae---and parameterizing them by their temperatures.
We adopted the effective temperature $T_\mathrm{eff}$ for each spectral type from those tabulated by \citet{Pecaut2013} and set the photosphere temperature $T_\mathrm{phot}$ to this value. 
We linearly interpolated within the grid of solar-metallicity stellar models of \citet{Siess2000} to determine masses and radii for early main sequence stars with these effective temperatures, with which we calculated surface gravities.

Typical starspot temperature contrasts vary as a function of stellar effective temperature, with larger temperature contrasts observed for hotter stars \citep[][and references therein]{Berdyugina2005}. 
We fitted a linear relation to the photosphere and spot temperatures of dwarfs presented in Table~5 of \citet{Berdyugina2005}, excluding the outliers of the solar penumbra and EK~Dra, and adopted the following relation for the spot temperature $T_{\text{spot}}$ as function of $T_{\text{phot}}$:
\begin{equation}
T_{\mathrm{spot}} = 0.418 \times T_{\mathrm{phot}} + 1620~\mathrm{K},
\label{eq:Tspot}
\end{equation}
in which both temperatures are given in Kelvin. 

Following \citet{Gondoin2008}, we adopted faculae temperatures of $T_{\text{fac}} = T_{\text{phot}} + 100$~K.
For comparison, \citet{Kobel2011} find an average contrast of 3.7\% in quiet Sun network magnetic elements with a broad range of about $-15\%$ to $+10\%$. 
A 3.7\% contrast on the Sun would correspond to roughly a 50~K increase over the photosphere, so we find the simple scaling relation that we adopt to be suitable.
We note, however, that this simple relation neglects the complex dependence of facular contrast on magnetic field strength and limb distance \citep{Norris2017}, which we save for consideration in a future work.

Table~\ref{tab:stars} lists the adopted surface gravities and photosphere, spot, and faculae temperatures for each spectral type. 
We note that the relation that we adopt for $T_\mathrm{spot}$ is determined using a stellar sample with effective temperatures between 3,300~K (M3) and 5,870~K (G1) \citep[][and references therein]{Berdyugina2005} and so may not hold for F dwarfs. 
However, in their study of rotation periods for main-sequence \textit{Kepler} targets with $T_\mathrm{eff} < 6500$~K, \citet{McQuillan2014} detect rotational variability for 4318 dwarfs with $T_\mathrm{eff} \in [5980, 6500]$~K (see their Table~1), corresponding to spectral types F5V to F9V.
They find a periodic detection fraction of 0.20 for stars with $T_\mathrm{eff} \in [6000, 6500]$~K, similar to the fractions for stars with $T_\mathrm{eff} \in [5000, 5500]$~K (0.27) and $T_\mathrm{eff} \in [5500, 6000]$~K (0.16).
We interpret this as evidence that the physical mechanism which drives rotational variability in G and K dwarfs, i.e., starspots and faculae, extends to stars as hot as F5 dwarfs.
Therefore, we adopt the scaling relation in Equation~\ref{eq:Tspot} for our full sample of spectral types.

For additional context, we briefly review in Section~\ref{sec:Fdwarfs} the literature on F dwarf photospheric features.

\begin{deluxetable}{ccccc}[!tbp]
\tablecaption{Adopted stellar parameters \label{tab:stars}}
\tablehead{
		   \colhead{Sp. Type}              &
		   \colhead{$T_{\text{phot}}$ (K)} &
		   \colhead{$T_{\text{spot}}$ (K)} &
           \colhead{$T_{\text{fac}}$ (K)}  &
           \colhead{log~$g$ (cgs)}                    
		  }
\startdata
F5V & 6510 & 4340 & 6610 & 4.32 \\
F6V & 6340 & 4270 & 6440 & 4.35 \\
F7V & 6240 & 4230 & 6340 & 4.36 \\
F8V & 6150 & 4190 & 6250 & 4.38 \\
F9V & 6040 & 4140 & 6140 & 4.40 \\
G0V & 5920 & 4090 & 6020 & 4.42 \\
G1V & 5880 & 4080 & 5980 & 4.43 \\
G2V & 5770 & 4030 & 5870 & 4.46 \\
G3V & 5720 & 4010 & 5820 & 4.47 \\
G4V & 5680 & 3990 & 5780 & 4.47 \\
G5V & 5660 & 3980 & 5760 & 4.48 \\
G6V & 5590 & 3960 & 5690 & 4.49 \\
G7V & 5530 & 3930 & 5630 & 4.50 \\
G8V & 5490 & 3910 & 5590 & 4.51 \\
G9V & 5340 & 3850 & 5440 & 4.54 \\
K0V & 5280 & 3830 & 5380 & 4.55 \\
K1V & 5170 & 3780 & 5270 & 4.56 \\
K2V & 5040 & 3730 & 5140 & 4.58 \\
K3V & 4840 & 3640 & 4940 & 4.61 \\
K4V & 4620 & 3550 & 4720 & 4.64 \\
K5V & 4450 & 3480 & 4550 & 4.67 \\
K6V & 4200 & 3370 & 4300 & 4.73 \\
K7V & 4050 & 3310 & 4150 & 4.78 \\
K8V & 3970 & 3280 & 4070 & 4.81 \\
K9V & 3880 & 3240 & 3980 & 4.85 \\
\enddata
\end{deluxetable}

\subsubsection{Note on F Dwarf Parameters}
\label{sec:Fdwarfs}

Going from hotter to cooler stars, chromospheric and coronal emission first occurs on the main sequence in the F dwarf stars, for which models of stellar structure also predict the onset of outer convection zones.
In particular, a sharp increase in the detection rate of stellar X-ray emission is seen at (B $-$ V) $\sim$ 0.3, coinciding with late-A-to-early-F main sequence stars \citep{Schmitt2001}.
While magnetic activity is clearly present, little is known about the morphology of the emergent magnetic flux regions in the photospheres of F dwarfs.
However, long-term monitoring programs in Ca~II H and K and Str{\"{o}}mgren photometry suggest a rather homogeneous spatial distribution of magnetic regions on F dwarfs.

\citet{Noyes1984} included 34 F stars (see their Table 1) in their study of rotation, convection, and activity on the main sequence, based on early results from intensive monitoring of the H and K lines to detect rotational modulation.
Of these, only 7 objects exhibited rotational modulation in their H and K lines while no periodic variability was seen in the remaining 27 stars, even though chromospheric activity in this sample was enhanced by an average factor of 1.5 compared to the Sun.
In their summary of the cycle properties of the stars in the Mt. Wilson Survey, \citet{Baliunas1995} included 40 F stars in their sample (see their Table 2).  
Of these, definitive cycle periods were measured in only 10 objects, which were all $\sim$ F5 or later.
Broad band photometric observations are consistent with the results from the H and K monitoring in the context of apparent departures from axisymmetric distributions of magnetic regions.  
In particular, \citet{Radick1982} found that photometric variability was not present in stars earlier than $\sim$~F7 at a detection limit of 0.5\%.  
Thus, F stars are characterized by a distinct lack of departures from axial symmetry of their surface distributions of magnetic flux.

The direct measurement of magnetic field properties on these stars has proven challenging because of their relatively more rapid rotation\footnote{This rapid rotation of F dwarfs can introduce equator-to-pole temperature gradients \citep{Deupree2011}, which represent a distinct source of stellar contamination that we do not consider in detail here.}
and the apparent absence of large-scale fields that typically give rise to spectrophotometric modulations.  
As discussed by \citet{Giampapa1984}, the relatively higher angular velocities of F dwarfs results in the generation of flux ropes at the base of the thin convection zone that are characterized by small spatial scales \citep{Schitt1983}.
Following magnetic flux rope dynamo generation, only minimal expansion of the emergent flux ropes is expected to occur.  
Thus, even though magnetic activity may be enhanced, large-scale inhomogeneities do not necessarily occur.  
The transition between this behavior in the limit of thin convection zones to solar-like, "thick" convection zones must occur at about F7V, since it is near this spectral type that photometric variability begins to appear, at least as documented in ground-based observations. 

Finally, we note that high activity in the form of X-ray or Ca~II emission can be present in F dwarfs even if only low-amplitude photometric variability is present. 
For example, \citet[][Table 4]{Pizzolato2003} show that saturated X-ray emission with $\log{L_\mathrm{X}} \sim$~30.1--30.3 occurs for F stars later than F5. 
\citet[][Figure 13]{Schroder2009} find enhanced $\log{R^{'}_\mathrm{HK}}$ values relative to the Sun in F dwarfs though with a declining envelope of values toward early F dwarfs, where convection zones are thinning out.  
Therefore, on one hand, low photometric variability among F dwarfs does not necessarily mean low magnetic activity.  
On the other hand, however, it is clear that activity is decreasing toward spectral types earlier than about F5--F7.

\subsection{Stellar Spectral Components}

For each spectral type, we generated spectra for the immaculate photosphere, spots, and faculae  using the PHOENIX stellar spectral model grid \citep{Husser2013}. 
We utilized models with solar metallicity ($\textrm{[Fe/H]} = 0.0$) and no $\alpha$-element enrichment ($[\alpha/\textrm{Fe}] = 0.0$). 
The PHOENIX grid provides high-resolution spectra covering 0.05--5.5~$\micron$ for effective temperatures in steps of 100~K for $T_\mathrm{eff} \in [2300, 7000]$~K and 200~K for $T_\mathrm{eff} \in [7000, 12000]$~K. 
Surface gravities are provided in steps of 0.5 for $\log{g} \in [0.0, 6.0]$. 
We linearly interpolated within the model grid in terms of temperature and surface gravity to produce component spectra with the parameters detailed in Table~\ref{tab:stars}.

\subsection{Rotational Variability Model}

\begin{figure*}[t]
\includegraphics[width=\linewidth]{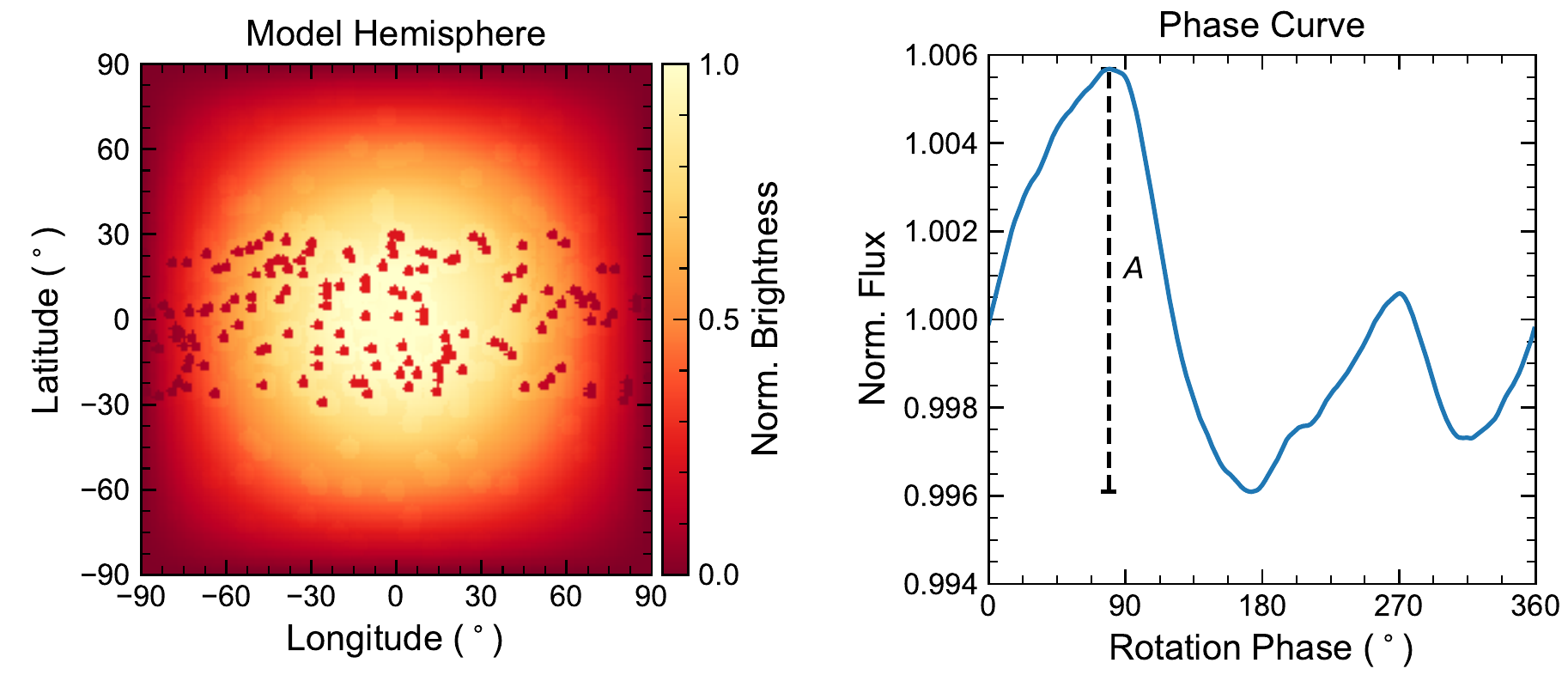}
\caption{An example of a model stellar photosphere and phase curve. 
The left panel shows one hemisphere of an example model photosphere with spots and facular regions after applying a double cosine weighting kernel. 
The right panel displays the phase curve produced by summing the hemispheric flux over one complete rotation of the model. 
The vertical dashed line illustrates the variability amplitude $A$, defined as the difference between the maximum and minimum normalized flux, which is $\sim 1$\% in this case. 
\label{fig:model}}
\end{figure*}

We employed the rotational modeling approach detailed in \ponet{} to investigate the range of photospheric heterogeneities consistent with observed variabilities. 
The approach involves iteratively adding active regions to a model photosphere and recording the peak-to-trough variability amplitude $A$ that results from rotating the model after each addition, as illustrated in Figure~\ref{fig:model}. 
Active regions are added at random coordinates until the model photosphere has the desired maximum spot coverage.
The entire process is repeated 100 times to build up statistics on the dependence of the variability amplitude on the spot covering fraction. 

We refer the reader to \ponet{} for a detailed description of the variability model and provide here the specifics for this work.
\added{However, one important assumption of this model bears repeating.
As in \ponet{}, we assume that the stellar rotation axis is aligned with the plane of the sky.
This assumption is good for most transiting exoplanet systems, since the presence of transits indicates a nearly edge-on planetary orbit, and obliquities between the stellar rotation axis and planetary orbital plane are generally $\lesssim 20 \degr$ \citep{Winn2017}.
The principal exceptions to this rule are hot stars with hot Jupiters, which tend to have high obliquities \citep{Schlaufman2010, Winn2010, Albrecht2012, Mazeh2015}.
The threshold stellar temperature above which these systems have a broader obliquity distribution is $6090^{+150}_{-110}$~K \citep{Dawson2014}, which roughly coincides with the ``Kraft break'' separating cool stars with convective envelopes from hot stars with radiative envelopes \citep{Kraft1967} and the boundary between F and G spectral types for our adopted parameters (Table~\ref{tab:stars}).
This suggests that determining spot and facula covering fractions from variability amplitudes for individual F-dwarf systems with hot Jupiters may require a more detailed treatment of the obliquity than the simple assumption that we make here.
Nonetheless, as we find the TLS spectral signals produced by F dwarfs to be relatively minor compared to those for later spectral types (see Section~\ref{sec:discussion}), we make this assumption for all models in this study and note that more detailed models may be required to investigate active region coverages in individual F-dwarf systems of interest or other notably oblique systems, such as the HAT-P-11 system \citep{Winn2010_HATP11b, Hirano2011_HATP11b, Yee2018_HATP11c}.}

As in \ponet{}, we used a model photosphere with a resolution of $180 \times 360$ pixels.
We simulated the immaculate photosphere, spots, and faculae by setting the pixel values to the flux of the component spectra integrated over the Kepler instrument response function\footnote{\url{http://keplergo.arc.nasa.gov/Instrumentation.shtml}}. 
This allows us to directly compare the rotational variabilities from our models to those reported by \citet{McQuillan2014}.
We fixed the spot size to $R_{spot} = 2 \degr$ so that each spot covered 400~ppm of a projected hemisphere (13 resolution elements), akin to large spot groups on the Sun \citep{Mandal2017}.  
While a detailed history of facular observations exists for the Sun \citep[e.g.,][]{Makarov1996, Shapiro2014}, little is known about the prevalence, distribution, and temperature contrasts for faculae on other stars\footnote{
Observations of transiting exoplanets, though, offer a promising probe of stellar photospheres that can shed light on this problem \citep[e.g.,][]{Dravins2017a, Dravins2017b, Dravins2018, Rackham2017, Espinoza2019}.}. 
Given this considerable uncertainty, we considered cases both with and without faculae. 
We refer to these hereafter as the \spotsfaculae{} and \spots{} cases, respectively. 
For the \spotsfaculae{} models, we added faculae at the 10:1 facula-to-spot area ratio observed for the active Sun \citep{Shapiro2014}, half of which were associated with spots and half of which were located independently, following the approach detailed in \ponet{}.

Spots on the Sun are found at active latitudes that vary predictably over the course of a solar cycle \citep{Maunder1904}, giving rise to the well-known butterfly diagram \citep{Maunder1922}. 
While individual sunspots can appear at latitudes as high as $\pm$ 40--50$\degr$, sunspot locations generally start around $28\degr$ from the equator at the beginning of a solar cycle and drift towards the equator over the course of a cycle \citep{Hathaway2011}. 
Spots on the K4 dwarf \object{HAT-P-11} have a mean latitude of $\approx 16\degr \pm 1 \degr$ and are generally found within $30\degr$ of the equator \citep{Morris2017}, which illustrates that the active latitudes observed on the Sun apply to at least some mid-K dwarfs as well. 
Faculae, on the other hand, are not confined to equatorial regions on the Sun; they can be found associated in spots or alone in polar regions \citep{Makarov1996}. 
Following these results, we restricted the locations of spots but not faculae to latitudes within $30\degr$ of the equator.

For each spectral type, we generated 100 model photospheres and added spots (and faculae) to each at randomly selected coordinates until we reached a full-disk spot covering fraction of 33\%.
This is the maximum spot coverage possible for our models, given the restriction on the spot latitudes.
From the set of 100 models, we calculated the mean and standard deviation of the variability amplitude as a function of spot covering fraction.

\added{Finally, we note that adopting a solar-like spot distribution in our models may cause us to underestimate the spot coverages and thus TLS signals from stars with notably different spot distributions.
Polar spots, for example, are commonly observed in Doppler images of rapidly rotating stars \citep[see][and references therein]{Strassmeier2009}.
If present on an exoplanet host star, such a spot configuration would contribute little to the rotational variability while producing a relatively large TLS signal.
While our intent here is to investigate typical active region coverages and TLS signals for Sun-like stars, studies of exoplanet host stars with suspected nonsolar active region distributions could benefit from other approaches such as stellar spectral decomposition \citep[e.g.,][]{Neff1995, Gully-Santiago2017} and simultaneous retrievals of stellar and planetary properties in transmission spectra \citep[e.g.,][]{Pinhas2018, Espinoza2019}.
We discuss some of these techniques in Section~\ref{sec:paths_forward}.}

\subsection{Variability as a Function of Spot Covering Fraction}

\begin{figure*}[!t]
\includegraphics[width=\linewidth]{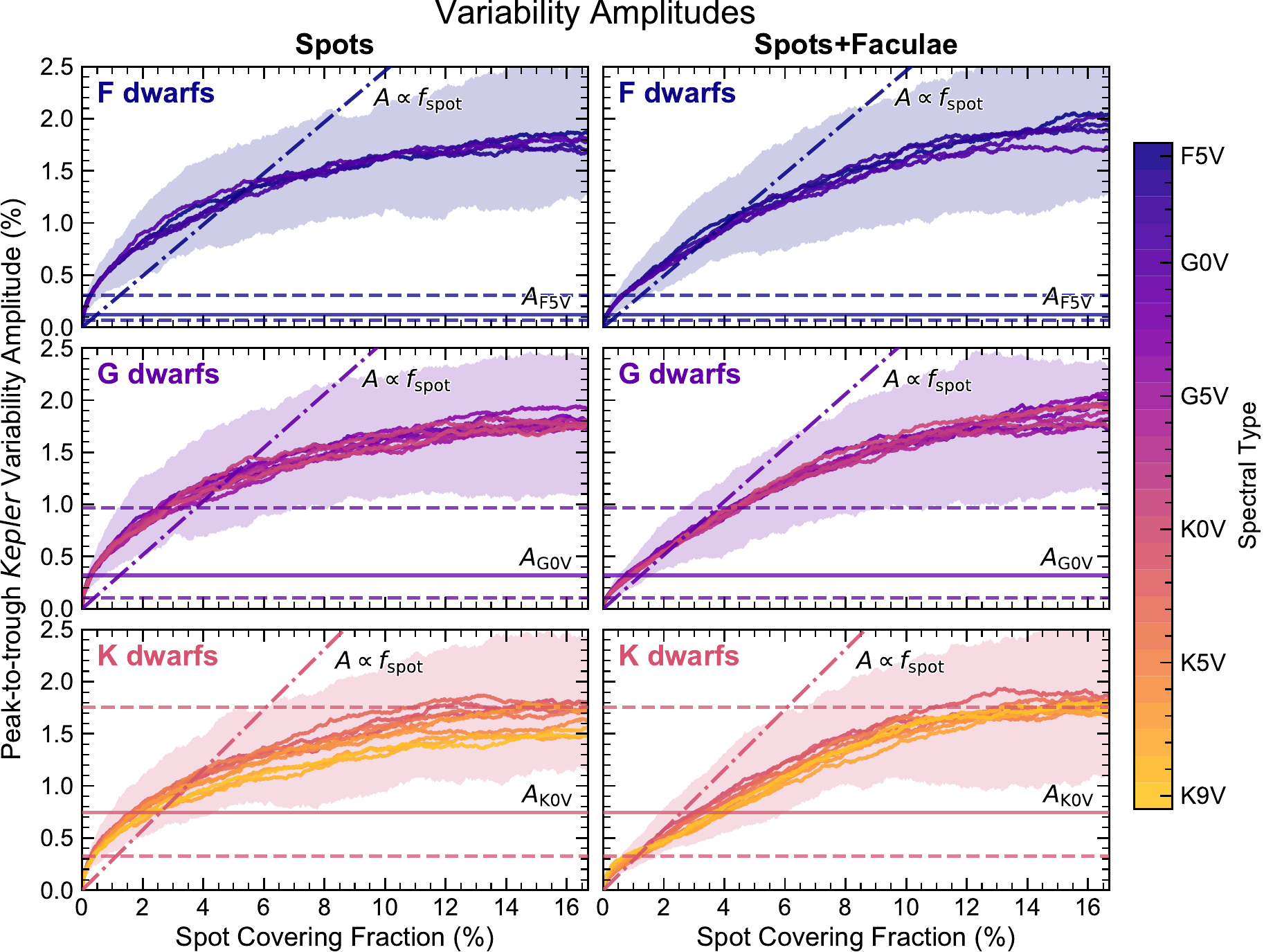}
\caption{Variability amplitudes as a function of spot covering fraction for \spots{} (left) and \spotsfaculae{} (right) models.
The top, middle, and bottom rows illustrate the results of models for F, G, and K main sequence spectral types, respectively. 
Solid curves give the mean variability as a function of spot covering fraction, color-coded by spectral type. 
For the earliest spectral type in each panel, the shaded region indicates the range encompassing 68\% of the model outcomes, which is comparable to the dispersion in model outcomes for all other spectral types. 
The dash-dotted line shows the expected scaling of spot covering fraction and variability for this spectral type, given its $Kepler$-band photosphere and spot fluxes and assuming a linear relation.
For all models, the linear relation is clearly not the appropriate scaling between the spot covering fraction and photometric variability, although it is a less poor approximation than in the case of M dwarfs \ponep{}.
Additionally, the horizontal lines show the median (solid line) and the 16th and 84th percentiles (dashed lines) of variability amplitudes for \textit{Kepler} dwarf stars for this spectral type \citep{McQuillan2014}. 
In each case, the variability grows asymptotically as a function of $f_\mathrm{spot}$ until reaching a maximum near 16.5\%, at which point half of the equatorial band is populated with spots and the photosphere is maximally heterogeneous. 
The dispersion in model outcomes leads to a range of spot covering fractions that correspond to a given amplitude. 
\label{fig:fspot_variability}}
\end{figure*}

We find that the variability amplitudes for each spectral type behave as a predictable function of the spot covering fraction. 
Variability amplitudes grow with increasing spot coverages until they reach a maximum near 16.5\%---at which point half of the photosphere within the allowed latitudes is covered in spots---and then decline as the spot coverages continue to increase to the maximum spot covering fraction and the equatorial band completely fills with spots. 
The behavior is roughly symmetric about spot coverages of 16.5\% and similar for both \spots{} and \spotsfaculae{} models.

As a result, variability amplitudes near the maximum amplitude correspond to a range of spot covering fractions, while smaller variability amplitudes correspond to two distinct spot covering fractions. 
Therefore, the relatively low variability amplitudes considered in this work (see Section~\ref{sec:reference_amplitude}) have both low- and high-spot-coverage solutions.
In this study, we are primarily interested in studying the extent of photospheric heterogeneities and the associated stellar contamination spectra for typical FGK stars.
While spot coverages of 33\% and higher have been identified for young and/or active stars such as \object{LkCa 4} \citep{Gully-Santiago2017}, we consider the turnover in variability amplitudes above 16.5\% spot coverage and the relatively low variabilities associated with nearly $33\%$ spot coverage to be artifacts of our model prescriptions.
Accordingly, we restrict our analysis to spot covering fractions below 16.5\%.
We note that a more realistic model would allow for a wider latitudinal spot distribution for very active stars (i.e., those with spot covering fractions 10\% and higher).

Figure~\ref{fig:fspot_variability} shows the variability amplitudes as a function of spot covering fraction for all spectral types that we consider.
The relationship between variability amplitudes and spot covering fractions appears similar for \spots{} and \spotsfaculae{} models. 
In both cases the variability amplitudes grow with a square-root-like dependence on the spot covering fraction. 
However, the variability amplitudes grow more slowly for low spot covering fractions in the \spotsfaculae{} models than in the \spots{} models, indicating that the presence of faculae tends to suppress the rotational variability.

These results contrast with those on M dwarfs, which show that the addition of faculae leads to large initial increases in variability amplitudes and larger amplitudes overall for the maximum spot covering fractions \ponep.
This difference owes to our model assumptions: 
We adopt a fixed temperature difference between the facula and immaculate photosphere components, which causes the facular contrast to decrease with increasing photosphere temperatures. 
The contrast is largest for M dwarfs and smallest for F dwarfs. 
Therefore, the inclusion of faculae in the models leads to large initial increases in variability for the coolest stars.
Additionally, for spot covering fractions above $\sim 10\%$, the \spotsfaculae{} models are completely covered by either spots or faculae, given the 10:1 facula-to-spot area ratio that we adopt. 
For M dwarfs, the contrast between spots and faculae is notably larger than that between spots and immaculate photosphere, which causes the overall larger variability amplitudes for the \spotsfaculae{} models. 
While for FGK stars, the spots/photosphere and spots/faculae contrasts are more comparable, resulting in the similar amplitudes for the \spots{} and \spotsfaculae{} models.

\added{
The apparent square-root-like dependence of the variability amplitude on the spot covering fraction can be understood as a consequence of the random longitudinal distribution of the spots.
For the case in which only spots contribute to the photospheric heterogeneity, the maximum brightness during a rotation will be at the longitude with the fewest spots, and the minimum brightness will be at the longitude with the most spots.
As the spots are distributed randomly in longitude, the expectation value for this difference for given number of spots $n$ will be on the order of $\sqrt{n}$, and the expectation value for the rotational variability amplitude will be on the order of $\alpha \Omega \sqrt{n}$, where $\alpha$ is the spot contrast ($1-F_\mathrm{spot}/F_\mathrm{phot}$)\footnote{In the \textit{Kepler} bandpass, the values of $\alpha$ for the models we use vary (nonmonotonically) from 0.86 for F0V to 0.73 for K9V.}
and $\Omega$ is the solid angle of the spot.
The dependence of $A$ on $f_\mathrm{spot}$ should therefore scale roughly as $\sqrt{f_\mathrm{spot}}$, since $n \sim f_\mathrm{spot}$.
For a given value of $f_\mathrm{spot}$, the relation should be steeper for larger spot contrasts and sizes.
Of course, the presence of faculae complicates this picture, as they can occur in association with spots or in isolated regions, and their brightness contribution thus weakens the relationship between spot coverage and longitudinal brightness.
The exact analytical dependence of $A$ on $f_\mathrm{spot}$ (and $f_\mathrm{fac}$) will depend on these parameters as well as inclination and limb-darkening effects \citep[as pointed out by][]{Jackson2012}; a full derivation of it is outside the scope of this analysis but could yield interesting insights in a future study.
For the present, we refer the reader to \citet{Jackson2013} for a more detailed discussion of the relation between variability amplitude and spot filling factor, typical size, and contrast.
}

Given the apparent square-root-like dependence of the variability amplitudes on the spot covering fraction, we fit via least squares a scaling relation of the form
\begin{equation}
A = C \times f_\mathrm{spot}^{0.5}
\label{eq:scaling_relation}
\end{equation}
to the variability amplitudes of the \spots{} and \spotsfaculae{} models for each spectral type, following \ponet. 
In this expression, $C$ is a scaling coefficient that depends on the properties of the active regions and determines the amplitude of the relation. 
Table~\ref{tab:scaling_coeffs} provides the fitted values of $C$ with uncertainties that reflect the 68\% dispersion in variability amplitudes illustrated by the shaded regions in Figure~\ref{fig:fspot_variability}.
These can be used to estimate spot covering fractions from observed variabilities of FGK main sequence stars.

\begin{deluxetable}{lcc}[tb]
\tablecaption{Scaling relation coefficients for variability models \label{tab:scaling_coeffs}}
\tablehead{
\colhead{Sp. Type} & \multicolumn{2}{c}{$C$} \\
\cline{2-3}
\colhead{} & \colhead{\spots{}} & \colhead{\spotsfaculae{}}
}
\startdata
F5V & $0.050 \pm 0.018$ & $0.051 \pm 0.020$ \\
F6V & $0.049 \pm 0.020$ & $0.050 \pm 0.020$ \\
F7V & $0.049 \pm 0.019$ & $0.051 \pm 0.019$ \\
F8V & $0.050 \pm 0.019$ & $0.049 \pm 0.020$ \\
F9V & $0.050 \pm 0.020$ & $0.047 \pm 0.018$ \\
G0V & $0.050 \pm 0.019$ & $0.047 \pm 0.018$ \\
G1V & $0.049 \pm 0.019$ & $0.050 \pm 0.019$ \\
G2V & $0.050 \pm 0.019$ & $0.050 \pm 0.019$ \\
G3V & $0.053 \pm 0.022$ & $0.050 \pm 0.019$ \\
G4V & $0.047 \pm 0.018$ & $0.047 \pm 0.018$ \\
G5V & $0.049 \pm 0.019$ & $0.048 \pm 0.018$ \\
G6V & $0.048 \pm 0.018$ & $0.048 \pm 0.018$ \\
G7V & $0.050 \pm 0.018$ & $0.049 \pm 0.019$ \\
G8V & $0.051 \pm 0.020$ & $0.047 \pm 0.018$ \\
G9V & $0.048 \pm 0.018$ & $0.051 \pm 0.019$ \\
K0V & $0.049 \pm 0.018$ & $0.048 \pm 0.019$ \\
K1V & $0.049 \pm 0.018$ & $0.050 \pm 0.018$ \\
K2V & $0.051 \pm 0.019$ & $0.047 \pm 0.019$ \\
K3V & $0.048 \pm 0.020$ & $0.047 \pm 0.018$ \\
K4V & $0.048 \pm 0.019$ & $0.046 \pm 0.018$ \\
K5V & $0.045 \pm 0.017$ & $0.045 \pm 0.016$ \\
K6V & $0.047 \pm 0.018$ & $0.045 \pm 0.018$ \\
K7V & $0.042 \pm 0.017$ & $0.043 \pm 0.017$ \\
K8V & $0.041 \pm 0.015$ & $0.046 \pm 0.017$ \\
K9V & $0.041 \pm 0.016$ & $0.046 \pm 0.017$ \\
\enddata
\end{deluxetable}

The values of $C$ show that the variability amplitudes are similar between the \spots{} and \spotsfaculae{} models, which illustrates that faculae do not strongly affect the rotational variability amplitudes. 
This finding is in agreement with results from the Sun, for which the signal from spots dominates the rotational brightness variations as viewed in the ecliptic plane \citep{Shapiro2016}\footnote{By contrast, \citet{Shapiro2016} also find that faculae dominate the long-term brightness variability on cycle time scales in the Sun and Sun-like stars for wavelengths shorter than 1.2~$\micron$, regardless of viewing inclination.}.

\subsection{Amplitude of Typical Activity Level} \label{sec:reference_amplitude}

\begin{figure}[!tbp]
\includegraphics[width=\linewidth]{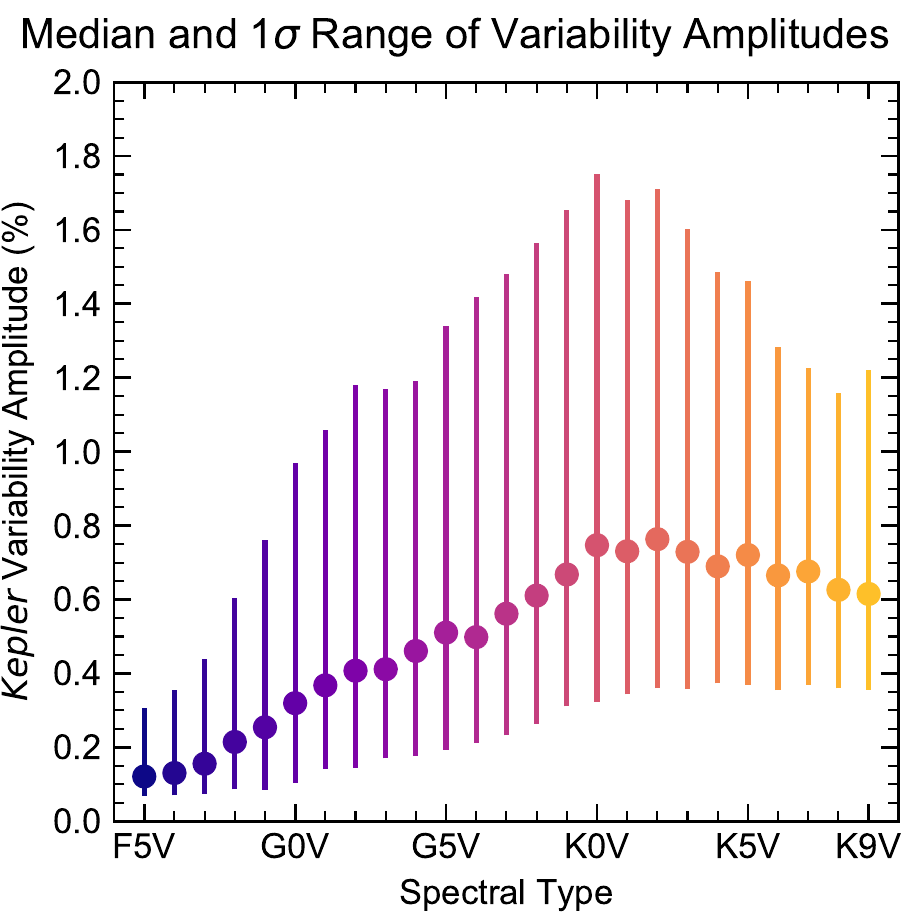}
\caption{Medians (points) and 68\% ranges (error bars) of \textit{Kepler} variability amplitudes by spectral type.
Data are summarized from \citet{McQuillan2014} \added{and color-coded by spectral type, following Figure~\ref{fig:fspot_variability}}.
See also Table~\ref{tab:variabilities}.
\label{fig:amplitudes}}
\end{figure}

In order to investigate typical levels of stellar contamination on transmission spectra, we must adopt a reference variability amplitude to use when estimating typical active region covering fractions. 
For the sun, disk passage of spots can decrease total solar irradiance by as much as $\sim 0.3\%$, while faculae can increase it by 0.1\% \citep{Willson1986}.
Turning to a wider sample, \citet{McQuillan2014} investigated periodic photometric variability amplitudes for the full \textit{Kepler} sample of main-sequence stars, building upon early analyses that focused on early subsets of the \textit{Kepler} data \citep{Basri2010, Basri2011}, specific spectral types \citep{McQuillan2013}, or exoplanet candidate host stars \citep{Walkowicz2013}.
For the full sample including spectral types F5V to M4V, they find a median amplitude, defined as the range between the 5th and 95th percentile of normalized flux, of $\sim 5600$~ppm or 0.56\%, though the typical amplitude varies with $T_\mathrm{eff}$.
In Table~\ref{tab:variabilities}, we summarize the data in their Table~1 for spectral types F5V--K9V separately, defining the spectral types by the outlined $T_\mathrm{eff}$ ranges.
We provide the median and $1\sigma$ (16th to 84th percentile) range of amplitudes for each spectral type bin.
These values are also illustrated in Figure~\ref{fig:amplitudes}.
Later spectral types show higher variability amplitudes on average.
Median variability amplitudes are highest and $1\sigma$ ranges widest for late G and early K dwarfs.

\begin{deluxetable}{ccrcc}[!tbp]
\tablecaption{Median and $1\sigma$ Range of Variability Amplitudes from \citet{McQuillan2014} by Spectral Type Bins \label{tab:variabilities}}
\tabletypesize{\footnotesize}
\tablehead{
  \colhead{Sp. Type} & \colhead{$T_\mathrm{eff}$ (K)} & $N_\mathrm{bin}$ & \multicolumn{2}{c}{Variability Amplitude} \\
  \cline{4-5}
  \colhead{} & \colhead{Range} & \colhead{} & \colhead{Median (\%)} & \colhead{$1\sigma$ (\%)}                    
		  }
\startdata
F5V & [6425, 6575) &  373 & 0.12 & [0.07, 0.31] \\
F6V & [6290, 6425) &  855 & 0.13 & [0.07, 0.36] \\
F7V & [6195, 6290) &  832 & 0.16 & [0.07, 0.44] \\
F8V & [6095, 6195) &  847 & 0.21 & [0.09, 0.61] \\
F9V & [5980, 6095) & 1411 & 0.25 & [0.08, 0.76] \\
G0V & [5900, 5980) & 1099 & 0.32 & [0.10, 0.97] \\
G1V & [5825, 5900) & 1106 & 0.37 & [0.14, 1.06] \\
G2V & [5745, 5825) & 1409 & 0.41 & [0.14, 1.18] \\
G3V & [5700, 5745) &  754 & 0.41 & [0.17, 1.17] \\
G4V & [5670, 5700) &  633 & 0.46 & [0.18, 1.19] \\
G5V & [5625, 5670) &  839 & 0.51 & [0.19, 1.34] \\
G6V & [5560, 5625) & 1379 & 0.50 & [0.21, 1.42] \\
G7V & [5510, 5560) & 1121 & 0.56 & [0.23, 1.48] \\
G8V & [5415, 5510) & 1926 & 0.61 & [0.26, 1.56] \\
G9V & [5310, 5415) & 2267 & 0.67 & [0.31, 1.66] \\
K0V & [5225, 5310) & 1703 & 0.75 & [0.32, 1.75] \\
K1V & [5105, 5225) & 2162 & 0.73 & [0.34, 1.68] \\
K2V & [4940, 5105) & 2737 & 0.76 & [0.36, 1.71] \\
K3V & [4730, 4940) & 2560 & 0.73 & [0.36, 1.60] \\
K4V & [4535, 4730) & 1550 & 0.69 & [0.37, 1.49] \\
K5V & [4325, 4535) & 1415 & 0.72 & [0.37, 1.46] \\
K6V & [4125, 4325) & 1793 & 0.67 & [0.35, 1.28] \\
K7V & [4010, 4125) &  799 & 0.68 & [0.37, 1.23] \\
K8V & [3925, 4010) &  449 & 0.63 & [0.36, 1.16] \\
K9V & [3865, 3925) &  272 & 0.62 & [0.36, 1.22] \\
\enddata
\end{deluxetable}

\replaced{We}
{We define a ``typically active'' star as one showing a rotational variability amplitude in the \textit{Kepler} bandpass equal to the median for its spectral type. Accordingly, we}
adopt the median amplitudes from Table~\ref{tab:variabilities} as the reference amplitudes $A_\mathrm{ref}$ that we use to determine the spot and facula covering fractions corresponding to the typical activity level for each spectral type.

\subsection{Spot and Facula Covering Fractions for Reference Amplitude} \label{sec:covering_fractions}

\begin{figure}[tbp]
\centering
\includegraphics[width=\linewidth]{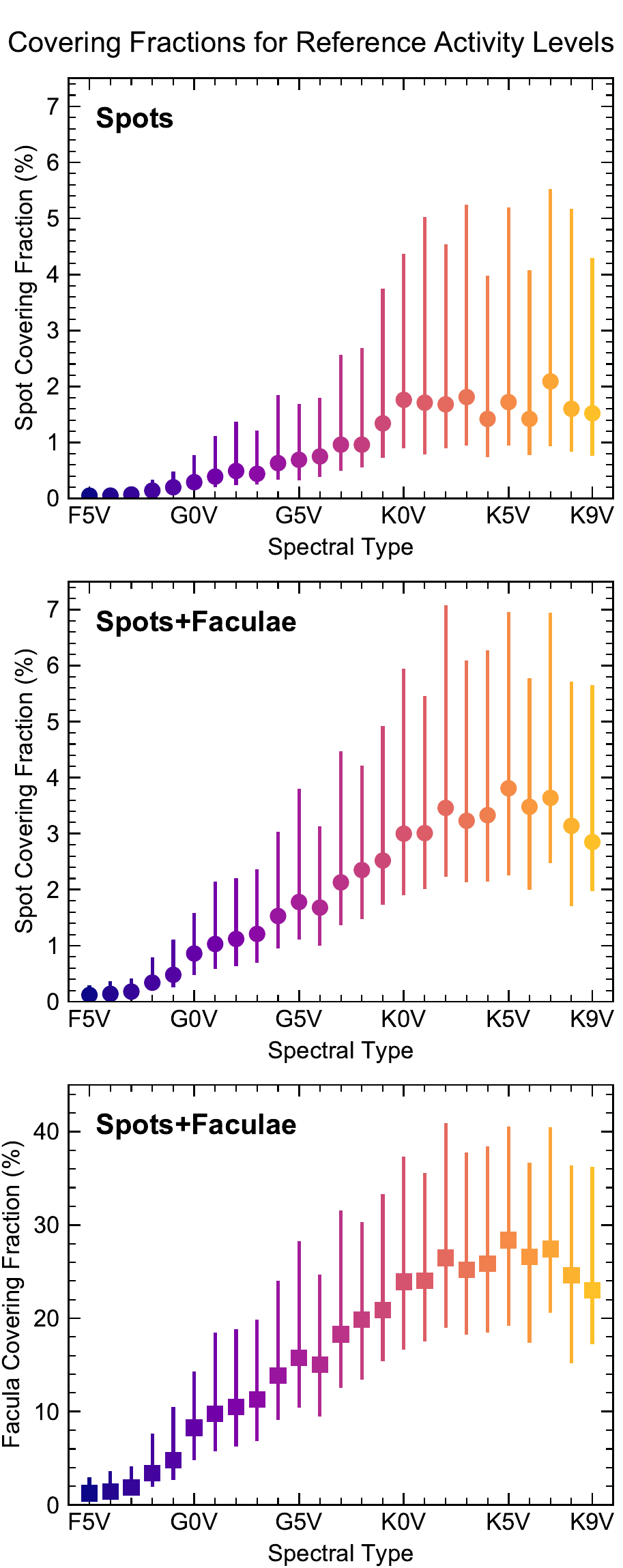}
\caption{Active region covering fractions corresponding to reference activity levels by spectral type.
Spot covering fractions for the \spots{} models are shown in the top panel.
For the \spotsfaculae{} models, spot (faculae) covering fractions are shown in the middle (bottom) panel.
\added{The data are color-coded by spectral type, following Figure~\ref{fig:fspot_variability}.}
See also Table~\ref{tab:filling_factors}.
\label{fig:filling_factors}}
\end{figure}

Table~\ref{tab:filling_factors} details and Figure~\ref{fig:filling_factors} illustrates the active region covering fractions corresponding to the reference amplitude for each spectral type. 
For each set of models, the mean covering fraction consistent with $A_\mathrm{ref}$ is given.
The uncertainties reflect the range of covering fractions that are consistent with $A_\mathrm{ref}$ for 68\% of the models.
In Figure~\ref{fig:fspot_variability}, this range is illustrated as the intersection of $A_\mathrm{ref}$ and the shaded $1\sigma$ envelopes for the variability amplitudes from the models.
Since $A$ and its $1\sigma$ envelope grow with a square-root-like dependence on $f_\mathrm{spot}$, this intersection produces uncertainties that are asymmetric and larger on the higher end.

\begin{deluxetable}{lccc}[tbp]
\tablecaption{Covering fractions for reference activity level as determined by variability models \label{tab:filling_factors}}
\tablehead{
\colhead{Sp. Type} & \colhead{\spots{}} & \multicolumn{2}{c}{\spotsfaculae{}} \\
\cline{2-2}
\cline{3-4}
\colhead{} & \colhead{$f_\mathrm{spot}$ (\%)} & 
\colhead{$f_\mathrm{spot}$ (\%)} & \colhead{$f_\mathrm{fac}$ (\%)}
}
\startdata
F5V & ${0.1}^{+0.1}_{-0.1}$ & ${0.1}^{+0.2}_{-0.1}$ & ${  1}^{  +2}_{  -1}$ \\
F6V & ${0.1}^{+0.1}_{-0.1}$ & ${0.1}^{+0.2}_{-0.1}$ & ${  1}^{  +2}_{  -1}$ \\
F7V & ${0.1}^{+0.1}_{-0.1}$ & ${0.2}^{+0.2}_{-0.1}$ & ${  2}^{  +2}_{  -1}$ \\
F8V & ${0.1}^{+0.2}_{-0.1}$ & ${0.3}^{+0.5}_{-0.1}$ & ${  3}^{  +4}_{  -1}$ \\
F9V & ${0.2}^{+0.3}_{-0.1}$ & ${0.5}^{+0.6}_{-0.2}$ & ${  5}^{  +6}_{  -2}$ \\
G0V & ${0.3}^{+0.5}_{-0.1}$ & ${0.9}^{+0.7}_{-0.4}$ & ${  8}^{  +6}_{  -3}$ \\
G1V & ${0.4}^{+0.7}_{-0.2}$ & ${1.0}^{+1.1}_{-0.5}$ & ${ 10}^{  +9}_{  -4}$ \\
G2V & ${0.5}^{+0.9}_{-0.2}$ & ${1.1}^{+1.1}_{-0.5}$ & ${ 10}^{  +8}_{  -4}$ \\
G3V & ${0.4}^{+0.8}_{-0.2}$ & ${1.2}^{+1.1}_{-0.5}$ & ${ 11}^{  +9}_{  -4}$ \\
G4V & ${0.6}^{+1.2}_{-0.3}$ & ${1.5}^{+1.5}_{-0.6}$ & ${ 14}^{ +10}_{  -5}$ \\
G5V & ${0.7}^{+1.0}_{-0.4}$ & ${1.8}^{+2.0}_{-0.7}$ & ${ 16}^{ +12}_{  -5}$ \\
G6V & ${0.8}^{+1.0}_{-0.4}$ & ${1.7}^{+1.5}_{-0.7}$ & ${ 15}^{ +10}_{  -6}$ \\
G7V & ${1.0}^{+1.6}_{-0.5}$ & ${2.1}^{+2.3}_{-0.8}$ & ${ 18}^{ +13}_{  -6}$ \\
G8V & ${1.0}^{+1.7}_{-0.4}$ & ${2.4}^{+1.9}_{-0.9}$ & ${ 20}^{ +10}_{  -6}$ \\
G9V & ${1.3}^{+2.4}_{-0.6}$ & ${2.5}^{+2.4}_{-0.8}$ & ${ 21}^{ +12}_{  -5}$ \\
K0V & ${1.8}^{+2.6}_{-0.9}$ & ${3.0}^{+2.9}_{-1.1}$ & ${ 24}^{ +13}_{  -7}$ \\
K1V & ${1.7}^{+3.3}_{-0.9}$ & ${3.0}^{+2.5}_{-1.0}$ & ${ 24}^{ +12}_{  -7}$ \\
K2V & ${1.7}^{+2.9}_{-0.8}$ & ${3.5}^{+3.6}_{-1.2}$ & ${ 26}^{ +14}_{  -7}$ \\
K3V & ${1.8}^{+3.4}_{-0.9}$ & ${3.2}^{+2.9}_{-1.1}$ & ${ 25}^{ +13}_{  -7}$ \\
K4V & ${1.4}^{+2.6}_{-0.7}$ & ${3.3}^{+2.9}_{-1.2}$ & ${ 26}^{ +13}_{  -7}$ \\
K5V & ${1.7}^{+3.5}_{-0.8}$ & ${3.8}^{+3.1}_{-1.6}$ & ${ 28}^{ +12}_{  -9}$ \\
K6V & ${1.4}^{+2.7}_{-0.7}$ & ${3.5}^{+2.3}_{-1.5}$ & ${ 27}^{ +10}_{  -9}$ \\
K7V & ${2.1}^{+3.4}_{-1.2}$ & ${3.6}^{+3.3}_{-1.2}$ & ${ 27}^{ +13}_{  -7}$ \\
K8V & ${1.6}^{+3.6}_{-0.8}$ & ${3.1}^{+2.6}_{-1.4}$ & ${ 25}^{ +12}_{  -9}$ \\
K9V & ${1.5}^{+2.8}_{-0.8}$ & ${2.9}^{+2.8}_{-0.9}$ & ${ 23}^{ +13}_{  -6}$ \\
\enddata
\end{deluxetable}

Considering the \spots{} models, we find the reference variability levels are consistent with mean spot covering fractions of $0.1\%$ to $2.1\%$ with $1\sigma$ ranges that are comparable to the means.
Generally, spot covering fractions are larger and their $1\sigma$ ranges are wider for later spectral types. 
For the \spotsfaculae{} models, spot covering fractions are systematically higher than those from the \spots{} models, though the values are consistent within their uncertainties. 
As expected, given our model assumptions, facula covering fractions are roughly 10 times larger than their spotted counterparts.
In effect, faculae dampen the rotational variability produced by spots at low activity levels, allowing larger covering fractions to be consistent with the same reference variability level. 
This effect is visible in Figure~\ref{fig:fspot_variability}, in which the relations between spot coverage and variability for the \spotsfaculae{} models are shifted to the right with respect to those of the \spots{} models.

\section{Stellar Contamination Analysis} \label{sec:contamination} 

With the active region covering fractions identified by the variability modeling, we can explore the typical levels of stellar contamination that we should expect for transmission spectra of exoplanets with FGK host stars.

\subsection{Model for Stellar Contamination Spectra}

We calculate the stellar contamination signal in the exoplanet transmission spectrum following the approach detailed in \ponet.
In short, we take the covering fractions identified in Section~\ref{sec:variability} and, assuming the exoplanet does not transit any active regions, calculate their effect on the observed transmission spectrum using the same stellar spectral components detailed above.

For the \spots{} case, the stellar contamination spectrum is given by 
\begin{equation}
\epsilon_\mathrm{\lambda, s} = 
	\frac{1}
	{1 - f_\mathrm{spot}(1 - \frac{S_\mathrm{\lambda, spot}}{S_\mathrm{\lambda, phot}})},
\label{eq:epsilon_s}
\end{equation}
in which $S_\mathrm{\lambda, spot}$ and $S_\mathrm{\lambda, phot}$ are the spot and immaculate photosphere spectra, respectively \citep[see also][]{McCullough2014, Rackham2017, Zellem2017}.
For the \spotsfaculae{} case, the stellar contamination spectrum is given by
\begin{equation}
\epsilon_\mathrm{\lambda, s+f} = 
	\frac{1}
	{1 - f_\mathrm{spot}(1 - \frac{S_\mathrm{\lambda, spot}}{S_\mathrm{\lambda, phot}})
       - f_\mathrm{fac}(1 - \frac{S_\mathrm{\lambda, fac}}{S_\mathrm{\lambda, phot}})},
\label{eq:epsilon_sf}
\end{equation}
in which $S_\mathrm{\lambda, fac}$ is the facula spectra and the remaining terms have the same meaning as above.

\added{
In general, for the case in which the planet occults a nominal emergent spectrum $S_{\lambda, 0}$ and $n$ other spectral components with covering fractions $f_{1}, f_{2}, \ldots, f_{n}$ are present elsewhere on the projected stellar disk, the stellar contamination spectrum is given by
\begin{equation}
\epsilon_\mathrm{\lambda, n} = 
	\frac{1}
	{1 - \sum_{i=1}^{n} f_{i}(1 - \frac{S_\mathrm{\lambda, i}}{S_\mathrm{\lambda, 0}})}.
\label{eq:epsilon_n}
\end{equation}
This expression is algebraically exact for a transit of disk of uniform intensity.
Since stellar disks actually display intensity profiles, it is an approximation to the true physical effect.
In greater detail, the emergent spectrum of the transit chord is most important near mid-transit, and its characteristics will therefore depend on the impact parameter of the transit.
Likewise, the contrast of the $n$th spectral component with the nominal spectral component $S_{\lambda, n}/S_{\lambda, 0}$ will depend on the wavelength-dependent intensity profile of the stellar disk, which can produce limb darkening or brightening or be relatively negligible at some wavelengths \citep[e.g.,][]{Claret2000}, and the location of the $n$th component.
We are interested here in examining the scale of the TLS effect for FGK stars generally, and so we ignore these higher-order effects, though in-depth studies involving precise observations of individual systems could benefit from considering them.
}

In 
\replaced{both}{all} 
cases, $\epsilon_{\lambda}$ represents a multiplicative change to the true transit depth (i.e., the square of the wavelength-dependent planet-to-star radius ratio $D_{\lambda} = (R_{\lambda, p}/R_{s})^{2}$) owing to the heterogeneity of the stellar photosphere.
This combines with the planetary signal to produce the observed transit depth:
\begin{equation}
D_\mathrm{\lambda, obs} = \epsilon_{\lambda} D_{\lambda}.
\label{eq:D_obs}
\end{equation}

As noted above, this calculation assumes that the light source illuminating the exoplanet atmosphere is described well by a single spectral component, $S_\mathrm{\lambda, phot}$. 
Of course, spots or faculae may be present within the transit chord as well in some cases.
This formalism still applies to these cases as long as the heterogeneities within the transit chord produce crossing events with amplitudes that are larger than the observation uncertainty, which allows them to be identified and taken into account \citep[e.g.,][]{Pont2008, Carter2011, Narita2013}.
In fact, crossing events are useful for understanding the stellar contamination of the transmission spectrum because they enable constraints on the sizes and contrasts of active regions (\citealt{Sanchis-Ojeda2011, Huitson2013, Mancini2013, Pont2013, Tregloan-Reed2013, Scandariato2017, Espinoza2019}; Bixel et al., submitted).
Still, active regions may be present within the transit chord with contrasts or sizes that do not allow them to be readily detected \citep{Mallonn2018}. 
More complicated models considering the distributions of heterogeneities both inside and outside the transit chord may be warranted by observations of more active host stars \citep[e.g.,][]{Zhang2018}, though that additional complication is beyond the scope of this work.

\subsection{Stellar Contamination Spectra}

\begin{figure*}[!htbp]
\includegraphics[width=\linewidth]{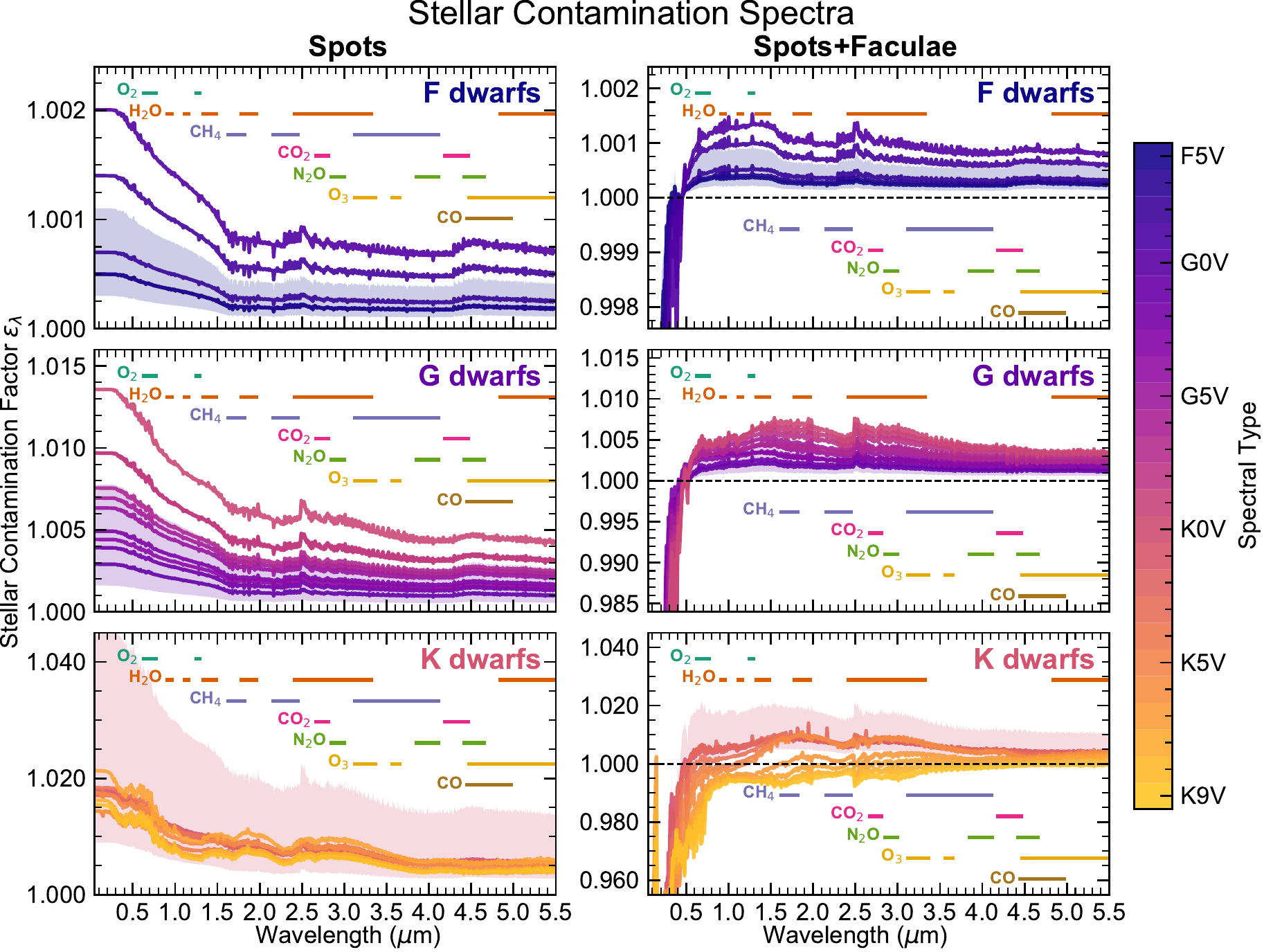}
\caption{Stellar contamination spectra for \spots{} (left) and \spotsfaculae{} (right) models. 
Contamination spectra for F, G, and K dwarfs are color-coded by spectral type and shown in the top, middle, and bottom panels, respectively. 
Solid lines indicate the contamination spectrum for the mean spot covering fraction consistent with the median variability amplitude for the spectral type (Table~\ref{tab:variabilities}).
The shaded regions illustrate the range of contamination spectra produced by spot covering fractions consistent with that same variability (see Table~\ref{tab:filling_factors}) for the earliest spectral type in each panel.
Wavelength bands for key molecular features in exoplanetary atmospheres are given. 
Note the different y-axis scales. 
\label{fig:contamination_spectra}}
\end{figure*}

Figure~\ref{fig:contamination_spectra} illustrates the stellar contamination spectra that correspond to the reference variability levels for each spectral type. 
For the \spots{} models, we find that unocculted spots consistent with $A_\mathrm{ref}$ increase transit depths at all wavelengths studied. 
The contamination spectra steadily increase with decreasing wavelengths for wavelengths shorter than $\sim 1.7 \micron$, producing apparent blueward slopes.
Late K-dwarf contamination spectra contain markedly more structure than their earlier spectral type counterparts.
In general, the scale of the contamination spectra increases for later spectral types.
The $1\sigma$ prediction intervals on the contamination spectra, dictated by the 68\% range of $f_\mathrm{spot}$ and illustrated by the shaded regions in Figure~\ref{fig:contamination_spectra}, are asymmetric, comparable to the absolute transit depth change (i.e., $|\epsilon_{\lambda}-1|$) on the upper end and roughly half that value on the lower end.

The contamination spectra for the \spotsfaculae{} models are generally similar to those of the \spots{} models but show strong differences at wavelengths shorter than $\sim 1.5\micron$.
As with the \spots{} model, the primary effect of the stellar contamination is to increase transit depths over most of the wavelengths studied.
However, owing to the presence of unocculted faculae, these spectra do not display the slopes seen at visual wavelengths with the \spots{} models.
Instead, they are relatively flat from the near-infrared (NIR) to wavelengths as short as $\sim 0.5\micron$ and then decrease sharply.
Later spectral types begin to show these decreases at longer wavelengths. 
For late K dwarfs, strong decreases in transit depth are possible across visual wavelengths; the effect of unocculted faculae can even overwhelm that of unocculted spots to produce decreases in transit depth over the full wavelength range studied.

Thus, considering exoplanet host stars with typical activity levels, we find that G dwarfs produce stellar contamination signals that are a factor a few larger than those of F dwarfs, while typically active K dwarfs produce signals that are more than an order of magnitude larger.
Unocculted faculae can partially cancel out the effect of unocculted spots at visual wavelengths and can lead to large decreases in transit depths at ultraviolet (UV) wavelengths.
We compare the scale of these stellar contamination signals to those of observational precisions and planetary atmospheric features in Section~\ref{sec:scale}.

Finally, we note that for all spectral types and stellar contamination models, the most significant effects are present at the shortest wavelengths.
This suggests that UV transit observations can therefore be used to place constraints on unocculted heterogeneities that affect transmission spectra more subtlety at longer wavelengths.
However, the stellar models used for this analysis lack chromospheres, which contribute significantly to emergent spectra at UV wavelengths, so a considerable level of uncertainty exists for the UV contamination spectra presented here.
Additionally, this picture is complicated by temporal variability of transit depths due to changing stellar activity levels \citep[e.g.,][]{Llama2015}.
We discuss the impact of chromospheres further in Section~\ref{sec:chromospheres}.

\subsubsection{Visual Stellar Contamination Spectra}

\begin{figure*}[!htbp]
\includegraphics[width=\linewidth]{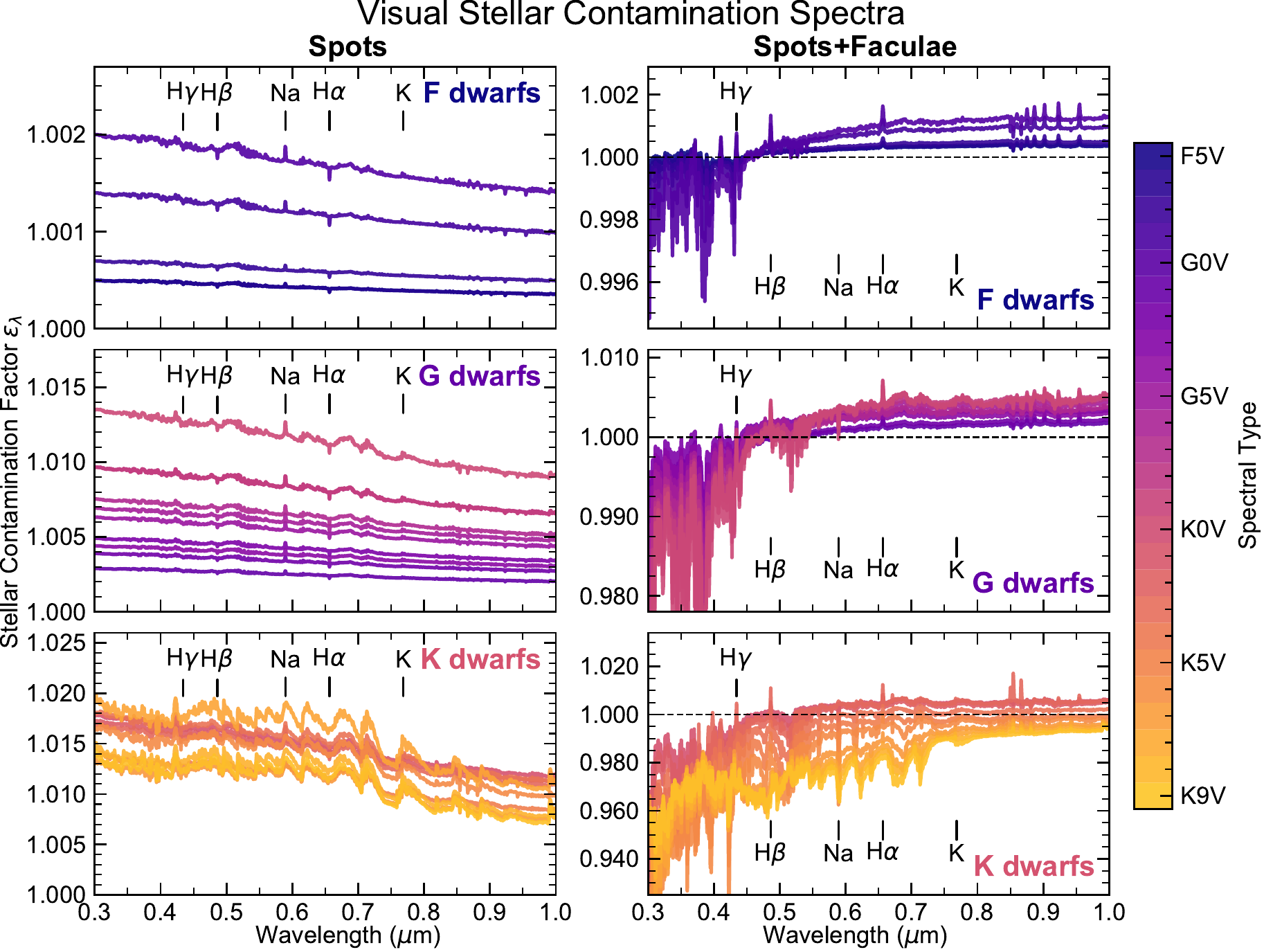}
\caption{Visual stellar contamination spectra for \spots{} (left) and \spotsfaculae{} (right) models. 
The figure elements are the same as those for Figure~\ref{fig:contamination_spectra}.
Prediction intervals on the spectra are suppressed for clarity. 
Key atomic absorption lines are indicated. 
Note the varying y-axis scales.
\label{fig:visual_spectra}}
\end{figure*}

Visual contamination features are of particular interest in the present study, given the availability of visual transmission spectra from both ground- and space-based facilities and the increased ability of stellar active regions to contaminate visual measurements. 
Figure~\ref{fig:visual_spectra} provides a closer look at the features in the stellar contamination spectra at visual wavelengths.
For the \spots{} models, the contamination spectra for F and G dwarfs show blueward slopes and few other spectral features.
Slight features are evident at wavelengths of atomic absorption for exoplanet atmospheres, namely Na~I, H$\alpha$, and K~I, which we explore further in Section~\ref{sec:discussion}.
K dwarfs, on the other hand, present more varied contamination spectra, with notable increases in transit depth around TiO molecular absorption features across visual wavelengths.
These features are strongest for later K dwarfs.

Turning to the \spotsfaculae{} models, we see in Figure~\ref{fig:visual_spectra} that the blueward slopes of the \spots{} models are muted by the presence of faculae, leading to contamination spectra that are generally flat for most visible wavelengths and decrease notably for wavelengths $\lessapprox 0.5~\micron$. 
Still, $\epsilon_{\lambda}$ remains systematically $>1$ for F and G stars at NIR wavelengths.
Late K dwarfs are again the exception, as their contamination spectra tend to decrease across all visible wavelengths and possess notably stronger spectral features than those of earlier spectral types.

\section{Discussion} \label{sec:discussion}

\subsection{Scale of the Stellar Contamination} \label{sec:scale}

\begin{deluxetable*}{lcccccc}[tb!]
\tablecaption{Mean values of stellar contamination spectra at UV, visual, and NIR wavelengths.  \label{tab:scales}}
\tablehead{
\colhead{Sp. Type} & \multicolumn{3}{c}{\spots{}} & \multicolumn{3}{c}{\spotsfaculae{}} \\
\cline{2-4}
\cline{5-7}
\colhead{} & \colhead{$\bar{\epsilon_\mathrm{UV}}$} & \colhead{$\bar{\epsilon_\mathrm{vis}}$} & \colhead{$\bar{\epsilon_\mathrm{NIR}}$} & 
\colhead{$\bar{\epsilon_\mathrm{UV}}$} & \colhead{$\bar{\epsilon_\mathrm{vis}}$} & \colhead{$\bar{\epsilon_\mathrm{NIR}}$}
}
\startdata
F5V & ${1.0005}^{+0.0006}_{-0.0002}$ & ${1.0004}^{+0.0005}_{-0.0002}$ & ${1.0002}^{+0.0003}_{-0.0001}$ & ${0.9960}^{-0.0054}_{+0.0017}$ & ${1.0002}^{+0.0004}_{-0.0001}$ & ${1.0003}^{+0.0004}_{-0.0001}$ \\
F6V & ${1.0005}^{+0.0005}_{-0.0001}$ & ${1.0004}^{+0.0004}_{-0.0001}$ & ${1.0002}^{+0.0002}_{-0.0001}$ & ${0.9947}^{-0.0077}_{+0.0022}$ & ${1.0002}^{+0.0004}_{-0.0001}$ & ${1.0003}^{+0.0005}_{-0.0001}$ \\
F7V & ${1.0007}^{+0.0009}_{-0.0002}$ & ${1.0006}^{+0.0008}_{-0.0002}$ & ${1.0003}^{+0.0004}_{-0.0001}$ & ${0.9930}^{-0.0082}_{+0.0031}$ & ${1.0003}^{+0.0004}_{-0.0001}$ & ${1.0004}^{+0.0005}_{-0.0002}$ \\
F8V & ${1.0014}^{+0.0020}_{-0.0006}$ & ${1.0012}^{+0.0017}_{-0.0005}$ & ${1.0006}^{+0.0008}_{-0.0002}$ & ${0.9868}^{-0.0151}_{+0.0056}$ & ${1.0006}^{+0.0010}_{-0.0003}$ & ${1.0007}^{+0.0010}_{-0.0003}$ \\
F9V & ${1.0020}^{+0.0028}_{-0.0008}$ & ${1.0017}^{+0.0024}_{-0.0007}$ & ${1.0008}^{+0.0012}_{-0.0003}$ & ${0.9797}^{-0.0215}_{+0.0088}$ & ${1.0007}^{+0.0013}_{-0.0003}$ & ${1.0010}^{+0.0014}_{-0.0005}$ \\
G0V & ${1.0029}^{+0.0048}_{-0.0013}$ & ${1.0024}^{+0.0041}_{-0.0011}$ & ${1.0012}^{+0.0020}_{-0.0005}$ & ${0.9640}^{-0.0228}_{+0.0143}$ & ${1.0004}^{+0.0011}_{-0.0003}$ & ${1.0015}^{+0.0015}_{-0.0007}$ \\
G1V & ${1.0039}^{+0.0073}_{-0.0019}$ & ${1.0033}^{+0.0061}_{-0.0016}$ & ${1.0016}^{+0.0030}_{-0.0008}$ & ${0.9575}^{-0.0312}_{+0.0166}$ & ${1.0006}^{+0.0023}_{-0.0005}$ & ${1.0018}^{+0.0025}_{-0.0009}$ \\
G2V & ${1.0049}^{+0.0089}_{-0.0025}$ & ${1.0041}^{+0.0075}_{-0.0021}$ & ${1.0020}^{+0.0036}_{-0.0010}$ & ${0.9332}^{-0.0366}_{+0.0233}$ & ${1.0022}^{+0.0035}_{-0.0011}$ & ${1.0025}^{+0.0028}_{-0.0011}$ \\
G3V & ${1.0044}^{+0.0078}_{-0.0019}$ & ${1.0037}^{+0.0066}_{-0.0016}$ & ${1.0018}^{+0.0032}_{-0.0008}$ & ${0.9301}^{-0.0370}_{+0.0240}$ & ${1.0016}^{+0.0032}_{-0.0009}$ & ${1.0024}^{+0.0028}_{-0.0011}$ \\
G4V & ${1.0063}^{+0.0124}_{-0.0029}$ & ${1.0053}^{+0.0105}_{-0.0025}$ & ${1.0026}^{+0.0051}_{-0.0012}$ & ${0.9202}^{-0.0398}_{+0.0234}$ & ${1.0018}^{+0.0045}_{-0.0011}$ & ${1.0030}^{+0.0038}_{-0.0013}$ \\
G5V & ${1.0069}^{+0.0102}_{-0.0037}$ & ${1.0058}^{+0.0086}_{-0.0031}$ & ${1.0028}^{+0.0042}_{-0.0015}$ & ${0.9111}^{-0.0447}_{+0.0249}$ & ${1.0022}^{+0.0070}_{-0.0013}$ & ${1.0036}^{+0.0054}_{-0.0015}$ \\
G6V & ${1.0075}^{+0.0107}_{-0.0037}$ & ${1.0063}^{+0.0090}_{-0.0031}$ & ${1.0031}^{+0.0043}_{-0.0015}$ & ${0.9099}^{-0.0386}_{+0.0277}$ & ${1.0014}^{+0.0040}_{-0.0010}$ & ${1.0033}^{+0.0036}_{-0.0015}$ \\
G7V & ${1.0096}^{+0.0164}_{-0.0047}$ & ${1.0081}^{+0.0138}_{-0.0040}$ & ${1.0039}^{+0.0066}_{-0.0019}$ & ${0.8877}^{-0.0474}_{+0.0274}$ & ${1.0019}^{+0.0081}_{-0.0014}$ & ${1.0042}^{+0.0064}_{-0.0017}$ \\
G8V & ${1.0096}^{+0.0178}_{-0.0040}$ & ${1.0081}^{+0.0149}_{-0.0034}$ & ${1.0039}^{+0.0072}_{-0.0017}$ & ${0.8664}^{-0.0378}_{+0.0311}$ & ${1.0021}^{+0.0064}_{-0.0017}$ & ${1.0047}^{+0.0051}_{-0.0020}$ \\
G9V & ${1.0134}^{+0.0251}_{-0.0063}$ & ${1.0113}^{+0.0209}_{-0.0052}$ & ${1.0055}^{+0.0101}_{-0.0026}$ & ${0.8044}^{-0.0314}_{+0.0181}$ & ${1.0018}^{+0.0086}_{-0.0015}$ & ${1.0050}^{+0.0068}_{-0.0019}$ \\
K0V & ${1.0177}^{+0.0274}_{-0.0088}$ & ${1.0148}^{+0.0228}_{-0.0074}$ & ${1.0072}^{+0.0109}_{-0.0036}$ & ${0.8001}^{-0.0316}_{+0.0247}$ & ${1.0022}^{+0.0119}_{-0.0022}$ & ${1.0061}^{+0.0087}_{-0.0026}$ \\
K1V & ${1.0172}^{+0.0349}_{-0.0093}$ & ${1.0143}^{+0.0289}_{-0.0078}$ & ${1.0070}^{+0.0139}_{-0.0038}$ & ${0.8080}^{-0.0353}_{+0.0277}$ & ${1.0004}^{+0.0086}_{-0.0016}$ & ${1.0058}^{+0.0070}_{-0.0023}$ \\
K2V & ${1.0168}^{+0.0299}_{-0.0079}$ & ${1.0140}^{+0.0247}_{-0.0065}$ & ${1.0068}^{+0.0119}_{-0.0032}$ & ${0.8596}^{-0.0415}_{+0.0305}$ & ${1.0010}^{+0.0150}_{-0.0024}$ & ${1.0063}^{+0.0107}_{-0.0028}$ \\
K3V & ${1.0180}^{+0.0360}_{-0.0086}$ & ${1.0149}^{+0.0297}_{-0.0072}$ & ${1.0074}^{+0.0144}_{-0.0035}$ & ${0.7950}^{-0.0498}_{+0.0377}$ & ${0.9961}^{+0.0083}_{-0.0007}$ & ${1.0057}^{+0.0081}_{-0.0024}$ \\
K4V & ${1.0139}^{+0.0260}_{-0.0067}$ & ${1.0115}^{+0.0214}_{-0.0055}$ & ${1.0057}^{+0.0105}_{-0.0028}$ & ${0.7955}^{-0.0497}_{+0.0401}$ & ${0.9916}^{+0.0066}_{+0.0005}$ & ${1.0051}^{+0.0080}_{-0.0024}$ \\
K5V & ${1.0167}^{+0.0354}_{-0.0075}$ & ${1.0137}^{+0.0288}_{-0.0062}$ & ${1.0069}^{+0.0142}_{-0.0031}$ & ${0.8646}^{-0.0337}_{+0.0351}$ & ${0.9882}^{+0.0068}_{+0.0013}$ & ${1.0050}^{+0.0083}_{-0.0029}$ \\
K6V & ${1.0136}^{+0.0264}_{-0.0063}$ & ${1.0110}^{+0.0212}_{-0.0051}$ & ${1.0055}^{+0.0105}_{-0.0025}$ & ${0.9473}^{-0.0099}_{+0.0150}$ & ${0.9850}^{+0.0018}_{+0.0031}$ & ${1.0014}^{+0.0044}_{-0.0015}$ \\
K7V & ${1.0201}^{+0.0349}_{-0.0113}$ & ${1.0159}^{+0.0274}_{-0.0089}$ & ${1.0078}^{+0.0132}_{-0.0044}$ & ${0.9407}^{-0.0122}_{+0.0116}$ & ${0.9826}^{+0.0034}_{+0.0024}$ & ${0.9996}^{+0.0057}_{-0.0009}$ \\
K8V & ${1.0153}^{+0.0358}_{-0.0074}$ & ${1.0120}^{+0.0279}_{-0.0058}$ & ${1.0058}^{+0.0132}_{-0.0028}$ & ${0.9406}^{-0.0165}_{+0.0195}$ & ${0.9813}^{-0.0010}_{+0.0054}$ & ${0.9980}^{+0.0029}_{-0.0001}$ \\
K9V & ${1.0145}^{+0.0276}_{-0.0073}$ & ${1.0113}^{+0.0213}_{-0.0057}$ & ${1.0053}^{+0.0099}_{-0.0027}$ & ${0.9312}^{-0.0250}_{+0.0147}$ & ${0.9795}^{-0.0033}_{+0.0039}$ & ${0.9968}^{+0.0022}_{+0.0003}$ \\
\enddata
\end{deluxetable*}

We first examine how the scale of the stellar contamination compares to that of observational precisions.
We consider UV, visual, and NIR wavelengths separately due to the distinct behaviors exhibited by the contamination spectra and the different observational approaches used to study these wavelength regimes.
Accordingly, we define the UV stellar contamination factor
$\bar{\epsilon_\mathrm{UV}}$,
the visual stellar contamination factor
$\bar{\epsilon_\mathrm{vis}}$,
and the NIR contamination factor
$\bar{\epsilon_\mathrm{NIR}}$
as the means of the contamination spectra for wavelength ranges 0.05--0.4~$\micron$, 0.4--0.9~$\micron$, and 0.95--5.5~$\micron$, respectively.
These values are provided in Table~\ref{tab:scales} for all FGK contamination spectra that we calculate, along with $1\sigma$ prediction intervals calculated by taking the means of their $1\sigma$ estimates (the shaded regions in Figure~\ref{fig:contamination_spectra}).

For the \spots{} models, the effects of stellar contamination are more pronounced at shorter wavelengths and for later spectral types.
For F dwarfs, we find the mean values of 
$\bar{\epsilon_\mathrm{UV}}$, $\bar{\epsilon_\mathrm{vis}}$, and $\bar{\epsilon_\mathrm{NIR}}$
are 1.0010, 1.0009, and 1.0004, respectively, all of which point to relatively minor increases in transit depths.
For G dwarfs, the corresponding means are 1.0069, 1.0058, and 1.0028, respectively, and for K dwarfs they are 1.0164, 1.0133, and 1.0066.

In the NIR, the scale of the contamination spectra for the \spotsfaculae{} models is comparable to that of the \spots{} models.
The mean value of $\bar{\epsilon_\mathrm{NIR}}$ is 1.0005, 1.0032, and 1.0030 for F, G, and K dwarfs, respectively.
By contrast, the corresponding means at visual wavelengths are smaller (1.0004, 1.0016, and 0.9908 for F, G, and K dwarfs, respectively) that those of the \spots{} models and point to absolute transit depth changes that are notably smaller.
This owes to the opposing signals of unocculted spots and faculae largely canceling out at visual wavelengths.
At UV wavelengths, however, the effects of unocculted faculae dominate and we find that the mean value of $\bar{\epsilon_\mathrm{UV}}$ is 0.9900, 0.9084, and 0.8683 for F, G, and K dwarfs, respectively.
In other words, unocculted faculae decrease transit depths in the UV, and the decreases are approximately 10\% of the transit depth for G and K dwarfs on average.

Whether these effects are detectable will depend on both observational precisions and the planet-to-star radius ratio of the system in question, as the stellar contamination signal scales with the nominal transit depth.
For comparison with observational precisions, we adopt 30~ppm as our fiducial detection threshold.
This is comparable to the typical transit depth uncertainty for current high-precision \textit{HST}/WFC3 transmission spectra observations \citep{Kreidberg2014} and systematic noise floors adopted by \citet{Greene2016} for NIRISS SOSS ($\lambda$ = 1--2.5~$\micron$; 20~ppm) and NIRCam grism ($\lambda$ = 2.5--5.0~$\micron$; 30~ppm) observations with 
\replaced{\textit{JWST}}
{the \textit{James Webb Space Telescope} (\textit{JWST})}.
For simplicity, we consider systems with a nominal transit depth of $D=1\%$, which corresponds to giant planets with radii ranging from $R=0.4~R_\mathrm{Jup}$ in the case of a K9V host star to $R=1.4~R_\mathrm{Jup}$ in the case of a F9V host.

Under these assumptions, stellar contamination would produce a 30~ppm feature and rise to the level of detectability when $|\epsilon_{\lambda} - 1| > 0.003$.
Therefore, considering the mean values tabulated in Table~\ref{tab:scales}, we find that for \spots{} models, the effects of unocculted spots for typically active FGK stars are detectable in the UV and visual for spectral types G1V and later and in the NIR for spectral types G6V and later.
For \spotsfaculae{} models, we find that the effects of unocculted heterogeneities are detectable in the UV for all spectral types F5V and later, while they are only detectable for K3V and later in the visual and G4V and later (excepting K6V and K7V) in the NIR.

To summarize, we find that unocculted heterogeneities in typically active G and K dwarfs can generally affect transmission spectra at levels relevant to current and near-future observational precisions.
The effects are less pronounced for F dwarfs, though the impact of unocculted faculae may be apparent at UV wavelengths for these stars.
While we focus on stars with typical activity levels here, we note that the stellar contamination signal obviously depends on the activity level of the star.
Therefore, more active stars can produce larger stellar contamination signals than we detail here, and these may be detectable for earlier spectral types.

How the scale of the stellar contamination compares to that of planetary transmission features will depend on the parameters of the exoplanet in question.
For the giant planets producing the nominal $D=1\%$ transit depths that we consider here, planetary transmission features are considerably larger than the 30~ppm threshold that we adopt \citep[e.g.,][]{Sing2016}.
Nonetheless, this analysis shows that for these planets, the stellar contamination signal of typically active FGK hosts can imprint on the transit depth at a scale that is detectable.
Therefore, we conclude that potential stellar contamination should be a consideration for all high-precision transmission spectroscopy studies of FGK-hosted exoplanets, particularly for observations with later host stars, more active hosts, and at shorter wavelengths.

\subsection{Visual Slopes} \label{sec:slopes}

The most prominent feature of the contamination spectra from the \spots{} models are the slopes produced at visual wavelengths.
They are of particular interest here because they can be potentially confused with scattering slopes originating in exoplanet atmospheres.

To quantify the scale of the visual slopes in the contamination spectra, we first define the average value of $\epsilon_{\lambda}$ in a wavelength bin $\Delta\lambda$ centered on some wavelength $\lambda_{0}$ as
\begin{equation}
\epsilon_\mathrm{avg}(\lambda_{0}, \Delta\lambda) = \frac{1}{\Delta\lambda} 
	\int_{\lambda_{0} - \Delta\lambda/2}^{\lambda_{0} + \Delta\lambda/2} \epsilon_{\lambda} d\lambda.
\label{eq:epsilon_avg}
\end{equation}
We then define the visual offset $\delta_\mathrm{vis}$ as
\begin{equation}
\delta_\mathrm{vis} = D [\epsilon_\mathrm{avg}(\lambda_{1}, \Delta\lambda)
                       - \epsilon_\mathrm{avg}(\lambda_{2}, \Delta\lambda)],
\label{eq:delta_visual}
\end{equation}
in which $D=1\%$, $\lambda_{1}=0.4~\micron$, $\lambda_{2}=0.9~\micron$, and $\Delta\lambda=0.1~\micron$.
Note that this formulation produces positive values for cases in which 
$\epsilon_{\lambda_{1}} > \epsilon_{\lambda_{2}}$, i.e. contamination spectra that increase towards shorter wavelengths.
For each spectral type and model framework (\spots{} and \spotsfaculae{}), we calculate $\delta_\mathrm{vis}$ from the mean contamination spectrum.
We also calculate $\delta_\mathrm{vis}$ for the upper and lower $1\sigma$ estimates for the contamination spectrum (i.e., the shaded regions in Figure~\ref{fig:contamination_spectra}), which we use to determine the $1\sigma$ prediction interval on $\delta_\mathrm{vis}$.

\begin{figure}[!tbp]
\includegraphics[width=\linewidth]{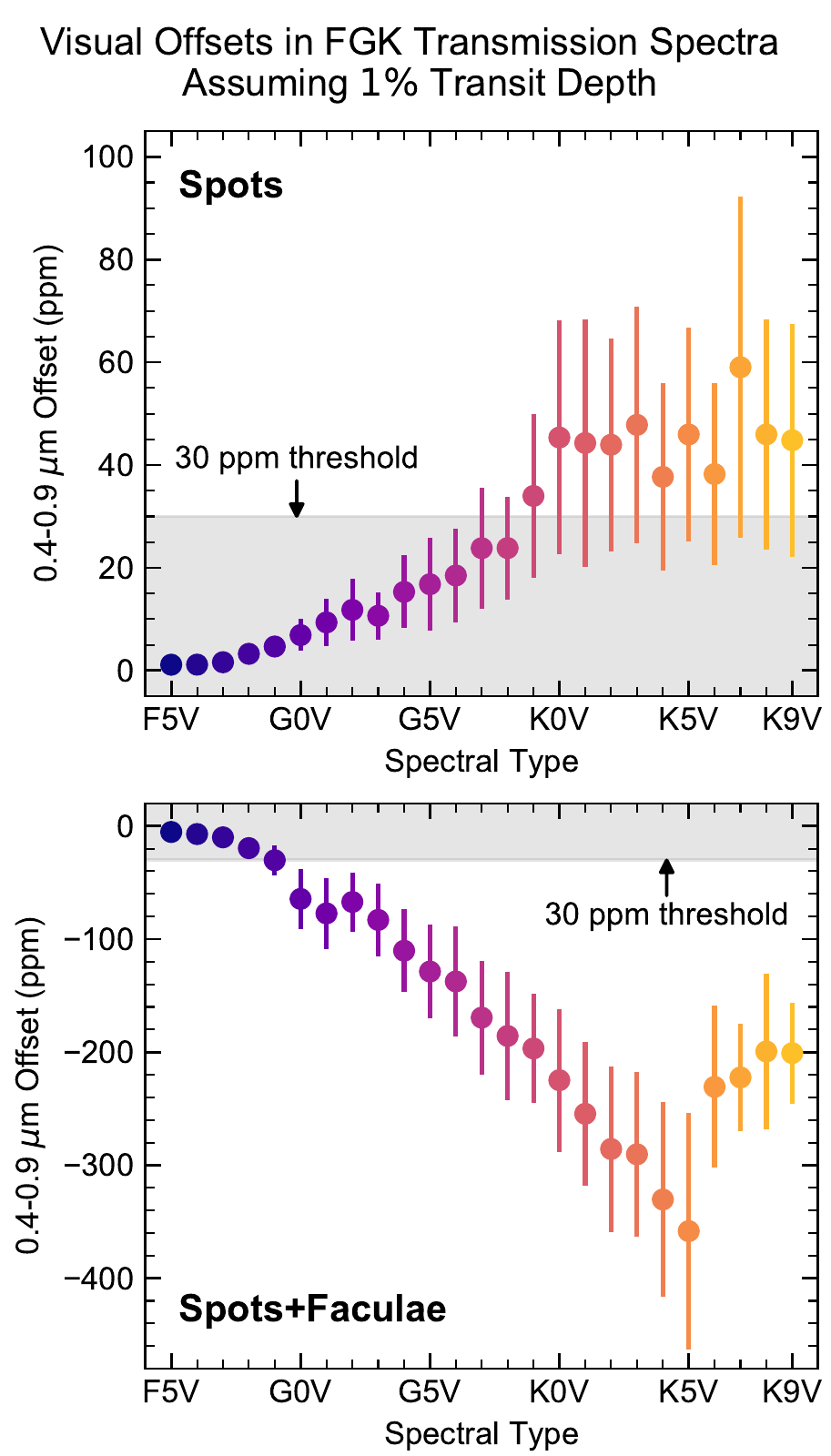}
\caption{Visual offsets in transmission spectra for \spots{} (top) and \spotsfaculae{} (bottom) models, assuming a nominal 1\% transit depth. 
For both model types, later spectral types produce larger visual offsets but with opposing signs.
The gray shaded region illustrates offsets that are below our adopted 30~ppm detection threshold.
\added{The data are color-coded by spectral type, following Figure~\ref{fig:fspot_variability}.}
Note the varying y-axis scales.
\label{fig:visual_offsets}}
\end{figure}

Figure~\ref{fig:visual_offsets} illustrates the visual offsets that we calculate for the \spots{} and \spotsfaculae{} models.
The \spots{} models produce positive visual offsets that increase in magnitude for later spectral types.
For spectral types G9V and later, $\delta_\mathrm{vis}$ is greater than the 30~ppm detection threshold, meaning that unocculted spots on a typically active G9V--K9V host star can produce detectable increases in transit depths across the visual.
However, these estimates are all consistent with the detection threshold at $1\sigma$.

The \spotsfaculae{} models, on the other hand, produce visual offsets that are negative and increase in magnitude more starkly for later spectral types.
We find that the absolute value of $\delta_\mathrm{vis}$ is greater than the detection threshold at $1\sigma$ confidence or higher for spectral types G0V and later.
Faculae on K5V host stars have the largest effect, producing visual offsets of $\delta_\mathrm{vis}=-350$~ppm at $3.4\sigma$ confidence.
These findings suggest that unocculted faculae can appreciably decrease visual transit depths in high-precision transmission spectra of exoplanets that orbit typically active G and K dwarfs.

This last point is interesting to consider in the context of the flat visual transmission spectra that are commonly observed for hot Jupiters \citep[e.g.,][]{Gibson2013, Huitson2017, Parviainen2018}.
These are counter to model predictions for clear atmospheres, which should show transit depths that increase at shorter wavelengths as a signature of Rayleigh scattering \citep{Seager2000, Fortney2010}.
Our results suggest that faculae can decrease visual transit depths at the level of a few hundreds of ppm, which is comparable to the precisions of current space-based \citep[e.g.,][]{Sing2016} and ground-based observations \citep[e.g.,][]{Espinoza2019, Nikolov2018}.
Therefore, it is possible that unocculted faculae could be counteracting signals from scattering slopes, making them at least in part responsible for the observed flat spectra.
This observation underscores the importance of atmospheric retrievals that consider both stellar and planetary signals in transmission spectra \citep{Espinoza2019, Pinhas2018}, which are discussed further in Section~\ref{sec:paths_forward}.

On the other hand, the effect of unocculted faculae depends on the value of $f_\mathrm{fac}$ for typically active G and K dwarfs, which we find could be between $\approx 8\%$ for early G dwarfs and $\approx 28\%$ for late K dwarfs (Table~\ref{tab:filling_factors}).
While He~I 10830~{\AA} equivalent width observations suggest that active F and G dwarfs can have active region filling factors of up to $\sim$~80--100\% \citep{Andretta2017}, the Sun at solar maximum only reaches a maximum annual average value of $f_\mathrm{fac} \approx 3\%$ \citep{Shapiro2014}, a factor of a few less than our estimates for early G dwarfs and roughly an order of magnitude less than our estimates for late K dwarfs.
Our approach relies on extrapolating the observed 10:1 facula-to-spot area ratio at solar maximum \citep{Shapiro2014} to higher activity levels.
However, this ratio may not hold for high activity levels generally.
Additionally, other stars may exhibit different facula-to-spot area ratios than the Sun does.
In this light, it possible that we have overestimated $f_\mathrm{fac}$ and therefore the effects of faculae on transmission spectra.
To complicate matters further, the Sun displays a time-dependence on the facula-to-spot area ratio throughout its activity cycle, with a facula-to-spot area ratio of 100:1 at solar minimum \citep{Shapiro2016}, primarily due to the absence of spots, and therefore we should expect other stars to do so as well.
Future efforts to constrain facular coverages for interesting exoplanet host stars generally and in a time-resolved way near transit observations could be useful in this respect.

\subsection{Atomic Absorption Features} \label{sec:atomic_features}

\begin{deluxetable}{cc}[tb!]
\tablecaption{Vacuum wavelengths used in analysis of transit depth line offsets \label{tab:lines}}
\tablehead{
		   \colhead{Feature}    &
           \colhead{Wavelength(s)}                    
		  }
\startdata
Na D doublet & 5894.570 \\
H$\alpha$    & 6564.665 \\
K~I doublet  & 7667.009, 7701.084 \\
\enddata
\end{deluxetable}

\begin{figure*}[!tbp]
\includegraphics[width=\linewidth]{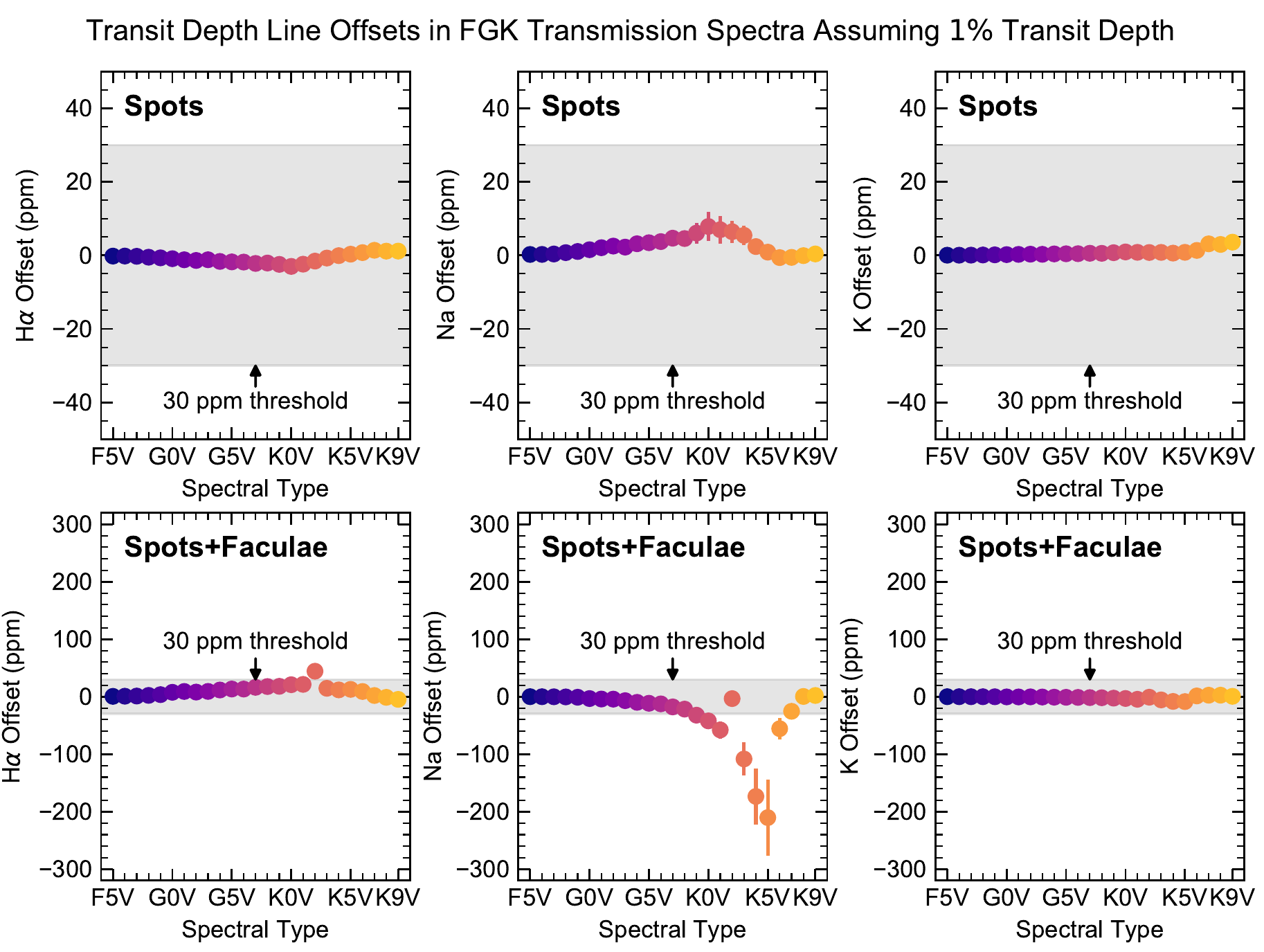}
\caption{Transit depth offsets in transmission spectra at H$_{\alpha}$ (left), the Na D doublet (middle), and the K doublet (right) for \spots{} (top) and \spotsfaculae{} (bottom) models, assuming a nominal 1\% transit depth. 
While \spots{} models for typically active FGK dwarfs do not produce detectable transit depth offsets at these line wavelengths, the results of the \spotsfaculae{} models suggest that detectable Na offsets are possible for K dwarfs and late G dwarfs.
The gray shaded region illustrates offsets that are below our adopted 30~ppm detection threshold.
The error bars indicate $1\sigma$ prediction intervals, which are generally smaller than the point size.
\added{The data are color-coded by spectral type, following Figure~\ref{fig:fspot_variability}.}
Note the varying y-axis scales.
\label{fig:line_offsets}}
\end{figure*}

The stellar contamination spectra for both \spots{} and \spotsfaculae{} models show distinct features at narrow atomic lines in the visual.
These include H$\alpha$, the Na D doublet, and the K~I doublet, all of which also produce prominent features in transmission spectra of giant exoplanets.
Broad absorption features from alkali metals point to cloud-free atmospheres and can be used to place constraints on their absolute abundances and in turn the atmospheric metallicity \citep{Nikolov2018}.
Alternatively, increases in transit depth only around the narrow cores of these lines point to the presence of clouds and hazes \citep[e.g.,][]{Sing2016}.
H$\alpha$ absorption features can be used to probe column densities and excitation temperatures in exoplanetary exospheres \citep{Jensen2012}.

To quantify the effects of these features on observations, we define the transit depth line offset $\delta_\mathrm{line}$ as 
\begin{equation}
\delta_\mathrm{line} = D [\epsilon_\mathrm{avg}(\lambda_\mathrm{line}, \Delta\lambda)
                        - \epsilon_\mathrm{cont}(\lambda_\mathrm{line}, \Delta\lambda)],
\label{eq:delta_line}
\end{equation}
in which the continuum value $\epsilon_\mathrm{cont}$ is calculated as
\begin{equation}
\epsilon_\mathrm{cont} = [\epsilon_\mathrm{avg}(\lambda_\mathrm{line}-1.5\Delta\lambda, \Delta\lambda)
                        + \epsilon_\mathrm{avg}(\lambda_\mathrm{line}+1.5\Delta\lambda, \Delta\lambda)]/2
\label{eq:epsilon_cont}
\end{equation}
and we set $D=1\%$ and $\Delta\lambda=20\textrm{\AA}$.

Table~\ref{tab:lines} lists the wavelengths used for the line offset analysis.
We obtained air wavelengths for these features from the NIST Handbook of Basic Atomic Spectroscopic Data\footnote{\url{https://www.nist.gov/pml/handbook-basic-atomic-spectroscopic-data}} and converted them to vacuum wavelengths following \citet{Birch1994}.
The individual lines of the sodium doublet are separated by only $6\textrm{\AA}$, so we used their average as the line wavelength.
As the individual lines of the potassium doublet are separated by more than $30\textrm{\AA}$, we calculated $\delta_\mathrm{line}$ for each line separately and, finding them to be comparable, report the mean.

Figure~\ref{fig:line_offsets} illustrates the line offsets that we calculate from the stellar contamination spectra and their $1\sigma$ prediction intervals.
None of the line offsets from the \spots{} models register above our 30~ppm detection threshold.
For the \spotsfaculae{} models, we find that no K offsets are detectable and neither are H$\alpha$ offsets, with the exception of an outlier at K2V.
\added{
These lines are relatively narrow in the model stellar spectra, and thus the line offsets are relatively insignificant when integrated over bandpasses relevant to low-resolution transmission spectroscopy (i.e., $\Delta\lambda=20\textrm{\AA}$).}
However, we find that Na offsets are detectable for spectral types G9V and later.
Interestingly, Na offsets generally trend smoothly towards more negative values for spectral types F0V to K4V before sharply turning around and decreasing to near zero for late K dwarfs.
Inspection of the stellar component spectra shows that the Na D doublet continuously broadens for spectral types from F0V to K9V.
For the latest K dwarfs, the Na D doublet becomes broader than $20\textrm{\AA}$, which complicates the determination of the continuum level.
Therefore, the turn-around in the Na offset seen for the latest K dwarfs is an artifact of our selection for $\Delta\lambda$ and not representative of a physical transition in the stellar atmospheres. 

The upshot of this analysis is that unocculted spots on typically active FGK dwarfs are not likely to produce detectable changes in transit depths around atomic features, though unocculted faculae can alter transit depths detectably around the Na D doublet.
However, the caveats discussed in Section~\ref{sec:slopes} regarding our prescriptions for modeling faculae apply here as well.

\subsection{Trends in Visual Features} \label{sec:trends}

\begin{figure*}[!htbp]
\includegraphics[width=\linewidth]{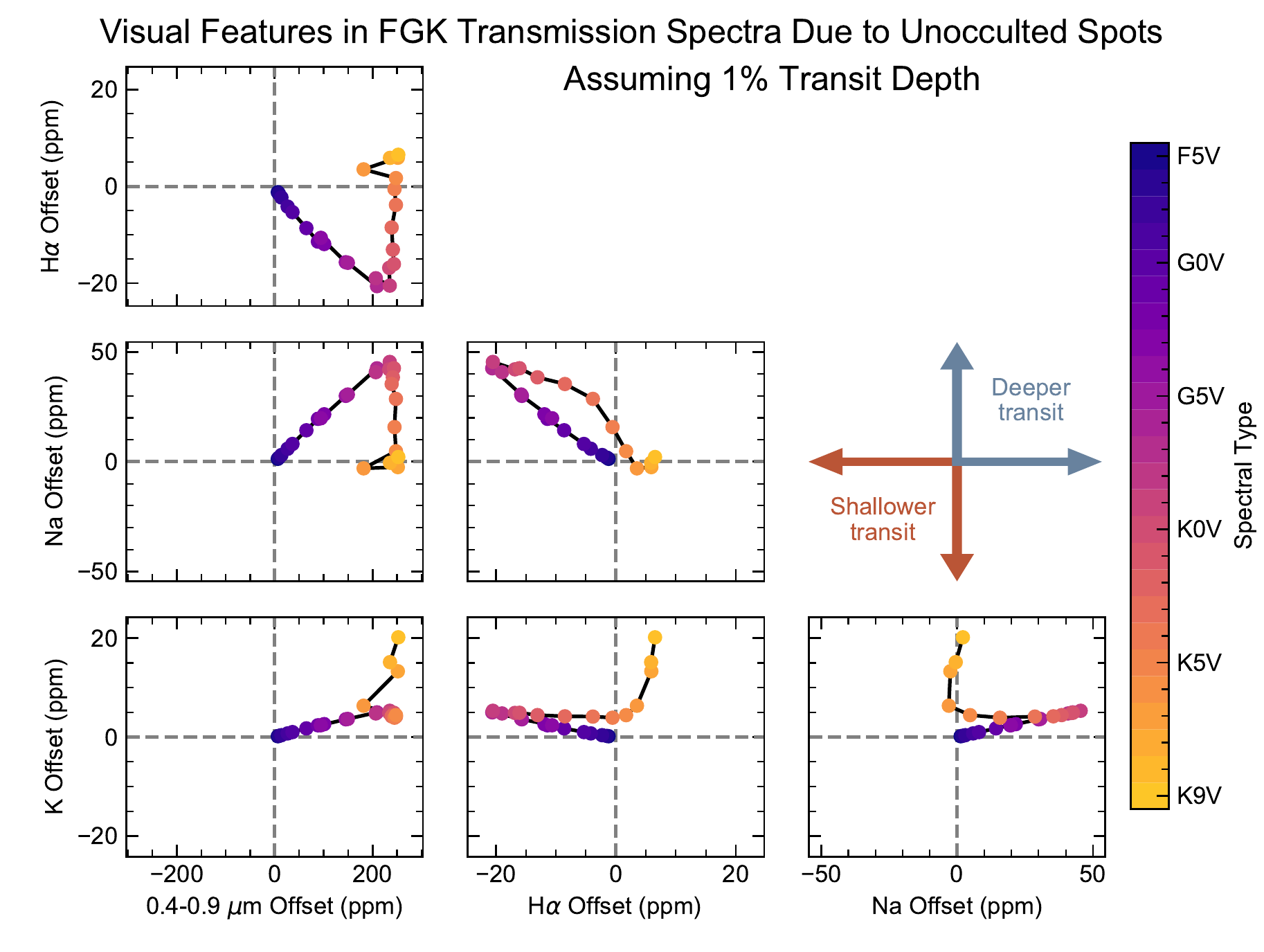}
\caption{Trends in transit depth changes in visual FGK contamination spectra features for \spots{} models. 
The magnitudes of stellar contamination features for FGK dwarfs generally grow with later spectral types.
They also trend in systematic ways in terms of their signs and relative strengths, which we suggest could be used to identify features with a stellar origin.
Positive (negative) values indicate deeper (shallower) transits.
Note the varying axis scales.
\label{fig:trends_s}}
\end{figure*}

\begin{figure*}[!htbp]
\includegraphics[width=\linewidth]{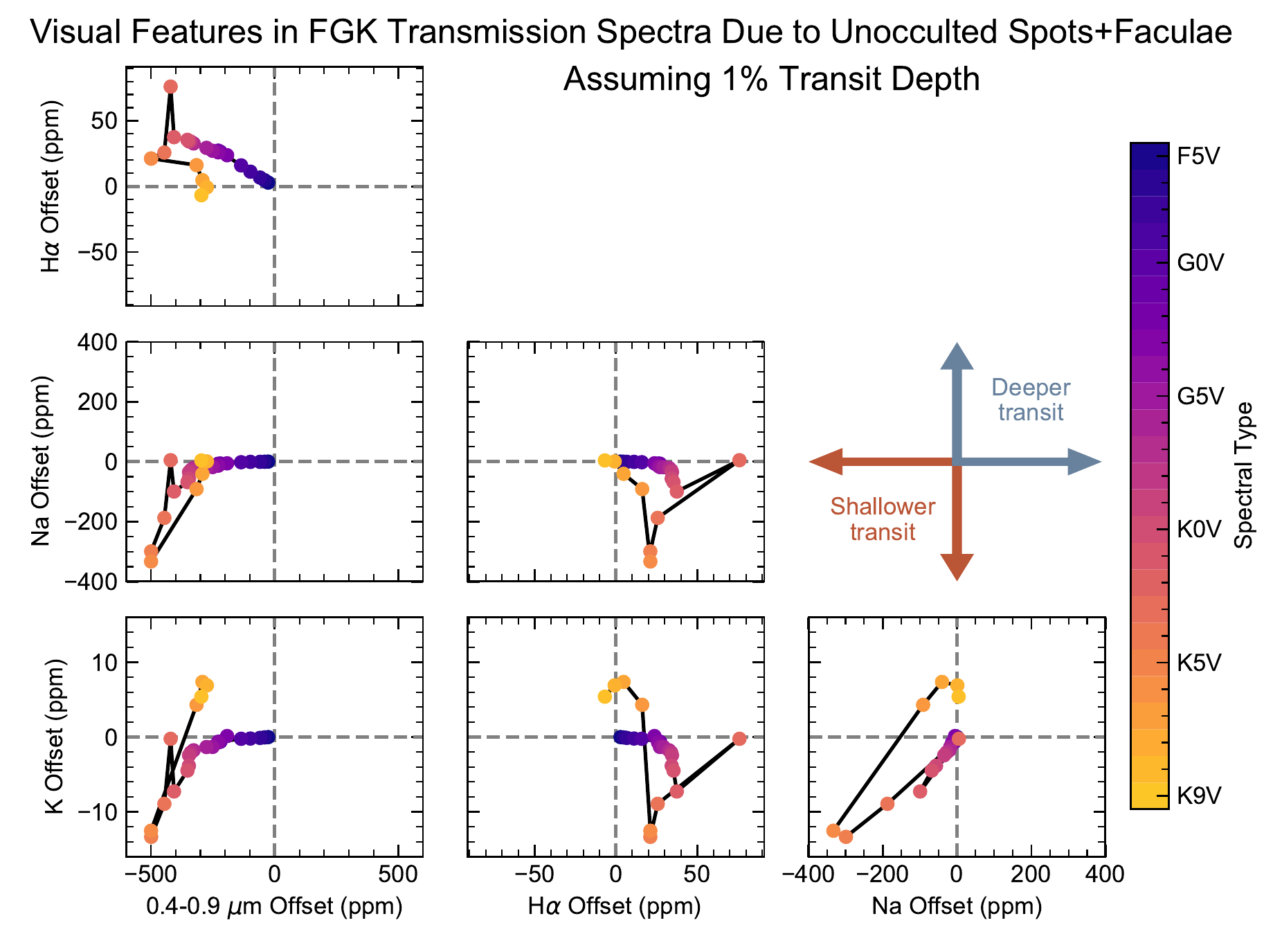}
\caption{Trends in transit depth changes in visual FGK contamination spectra features for \spotsfaculae{} models. 
As with the \spots{} models, the magnitudes of the stellar contamination features grow for later spectral types, though the signs of the features differ, which indicates that the effects of unocculted faculae dominate at visual wavelengths.
Positive (negative) values indicate deeper (shallower) transits.
Note the varying axis scales.
\label{fig:trends_sf}}
\end{figure*}

The analysis in the previous sections shows that, with a few exceptions, visual stellar contamination features are generally not detectable in transmission spectra of exoplanets hosted by typically active FGK dwarfs.
However, more active host stars may still be problematic.
To investigate this, we repeated the analysis presented in Section~\ref{sec:contamination},
\replaced{using}
{defining an ``active star'' as one showing a rotational variability amplitude in the \textit{Kepler} bandpass equal to the 84\% percentile for its spectral type. Accordingly, we use as the reference amplitudes for this ``active case'' the $1\sigma$ upper limits on the variability amplitudes (i.e., the 84\% percentiles) from Table~\ref{tab:variabilities}}.
\replaced{For this ``active case'',}
{In this case,}
the active region covering fractions that correspond to the reference amplitude are a factor of a few higher than for the nominal case.
Specifically, in the \spots{} case, $f_\mathrm{spot}$ is 7 times larger on average, while in the \spotsfaculae{} case, $f_\mathrm{spot}$ and $f_\mathrm{fac}$ are 4 and 3 times larger, respectively.

These larger covering fractions produce larger stellar contamination signals, making more features detectable above our adopted threshold.
In general, the offsets trend with spectral type in the same manner as shown in Figures~\ref{fig:visual_offsets} and \ref{fig:line_offsets} but the scales of the offsets are exaggerated.
For the \spots{} models, positive visual offsets are larger than 30~ppm for spectral types F9V and later and reach a peak of 254~ppm at spectral type K5V.
Additionally, positive Na offsets are $>30$~ppm for spectral types G4V--K2V.
For the \spotsfaculae{} models, the negative visual offsets are $>30$~ppm in magnitude for all spectra types and reach a peak value of $-483$ at spectral type K5V.
Positive H$\alpha$ offsets and negative Na offsets are detectable for late G to mid K dwarfs.
K offsets are smaller than our adopted threshold for all spectral types but begin to increase for late K dwarfs.

The magnitudes of these offsets, particularly with respect to the visual slope and the Na line offset, are such that they could be confused with features originating the the atmospheres of transiting exoplanets.
However, these features trend with each other and with spectral type in systematic ways.
These trends can be used to identify features with a stellar origin and disentangle them from planetary ones.

Figure~\ref{fig:trends_s} illustrates these trends for the \spots{} models in the active case. 
Generally, all offsets are near-zero for the earliest spectral type and increase for later spectral types.
The largest offset overall is the visual (0.4--0.9~$\micron$) offset, and the largest line offset is that of Na.
Starting with F5V, these both increase for later spectral types, reaching maxima around late G dwarfs, after which the visual offsets remain roughly the same, while the Na offsets decrease.
The signs and relative magnitude of these features could point to a stellar origin for features observed in transmission spectra, particularly for exoplanets hosted by late G or K dwarfs.

Figure~\ref{fig:trends_sf} illustrates the observed trends in offsets for the \spotsfaculae{} models.
Compared to the \spots{} models, the offsets have larger magnitudes and the trends have the opposite signs, due to the ability of unocculted faculae to dominate the visual slope and line offsets.

\replaced{We}
{In general, identifying these trends in observations remains a challenge, given the current state-of-the-art precision.
Of the trends illustrated in Figures~\ref{fig:trends_s} and \ref{fig:trends_sf}, only that between the Na line offset and the visual slope in Figure~\ref{fig:trends_sf} produces changes in both features well above the adopted 30~ppm detection threshold (and, essentially, only for K dwarfs).
This points to a potential limitation of the usefulness of these diagnostics.
However, these trends will be more evident in stars that are more active than our ``active case'' (recalling that we define an ``active star'' as one with a rotational variability only $1\sigma$ above the median), and future observational precisions may very well push below the 30~ppm threshold.
For these reasons, we}
point out these trends so that they may be of use in disentangling stellar and planetary features in transmission spectra in future studies.

\subsection{Molecular Absorption Features} \label{sec:molecular_features}

\begin{figure*}[!tbp]
\includegraphics[width=\linewidth]{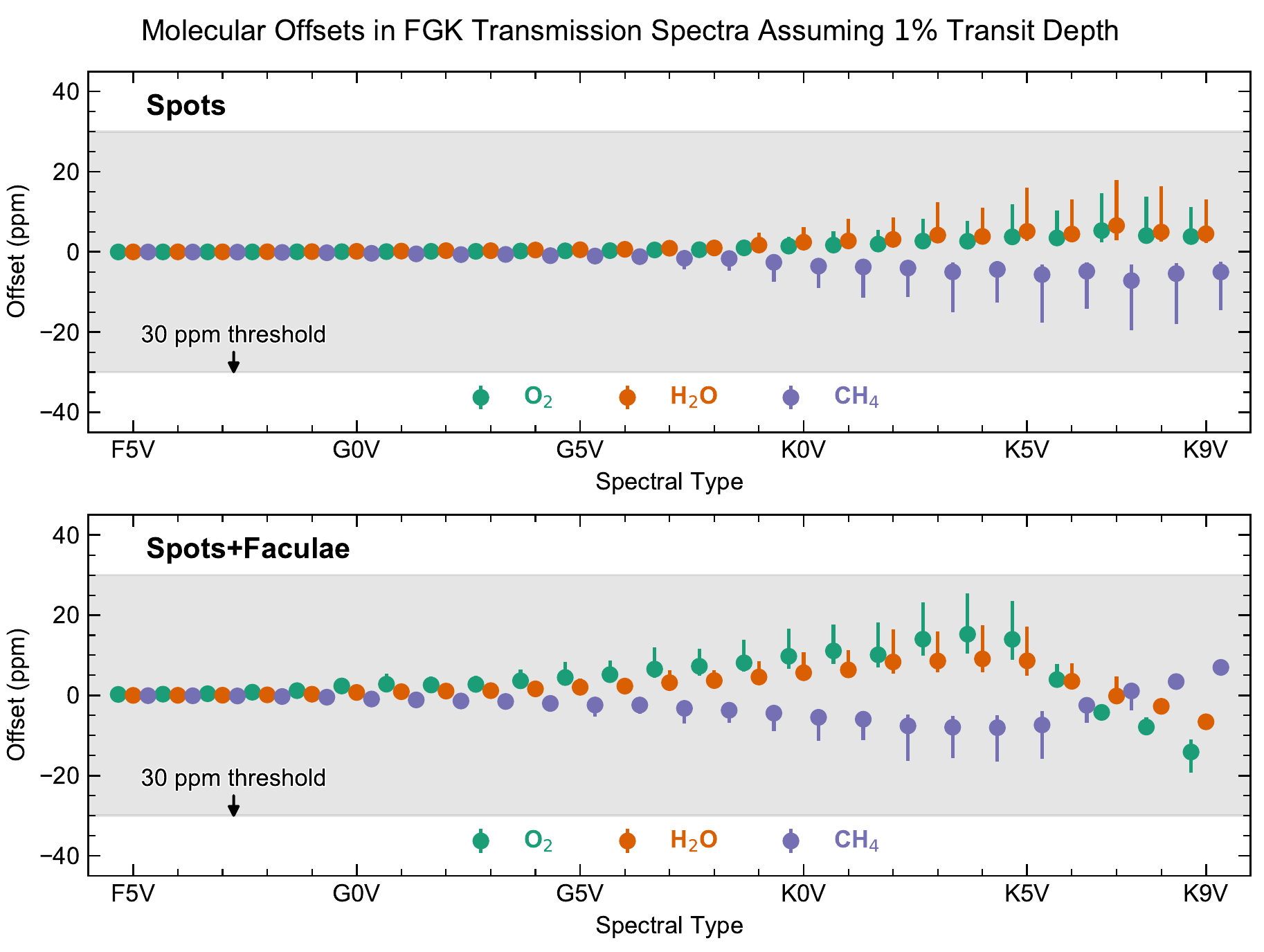}
\caption{Offsets in transmission spectra at wavelengths of interest for O$_2$, H$_{2}$O, and CH$_{4}$ for \spots{} (top) and \spotsfaculae{} (bottom) models, assuming a nominal 1\% transit depth. 
While offsets generally grow in magnitude for later spectral types, none are above our adopted 30~ppm detection threshold.
The gray shaded region illustrates offsets that are below our adopted 30~ppm detection threshold.
\label{fig:molecular_offsets}}
\end{figure*}

In addition to enabling studies of atomic absorption features, transmission spectra are useful probes of molecular absorption bands in exoplanet atmospheres.
The contamination spectra plotted in Figure~\ref{fig:contamination_spectra} show broad features owing to changes in molecular opacities between the immaculate photosphere and stellar active regions.
Figure~\ref{fig:contamination_spectra} also illustrates wavelengths of interest for some potentially detectable molecules in exoplanet atmospheres, including CH$_{4}$, CO, CO$_{2}$, H$_{2}$O, N$_{2}$O, O$_{2}$, and O$_{3}$.
If the values of stellar contamination spectra within these bands differ systematically from those of adjacent wavelengths, the stellar signal could mimic or mask exoplanetary molecular features in transmission spectra.

We investigate this possibility quantitatively following a similar approach to the analysis of atomic absorption features detailed in Section~\ref{sec:atomic_features}.
We define the transit depth band offset as
\begin{equation}
\delta_\mathrm{band} = D [\epsilon_\mathrm{avg}(\lambda_\mathrm{band}, \Delta\lambda_\mathrm{band})
                        - \epsilon_\mathrm{cont}(\lambda_\mathrm{line}, \Delta\lambda_\mathrm{band}, \Delta\lambda)],
\label{eq:delta_band}
\end{equation}
in which $\lambda_\mathrm{band}$ is the central wavelength of the molecular band, $\Delta\lambda_\mathrm{band}$ is its width, and we set $D=1\%$ as before.
In this case, the continuum value $\epsilon_\mathrm{cont}$ is calculated as
\begin{equation}
\epsilon_\mathrm{cont} = [\epsilon_\mathrm{avg}(\lambda_0, \Delta\lambda) \\
                        + \epsilon_\mathrm{avg}(\lambda_1, \Delta\lambda)]/2
\label{eq:epsilon_cont2}
\end{equation}
in which $\lambda_0 = \lambda_\mathrm{band}-\Delta\lambda_\mathrm{band}-0.5\Delta\lambda$,
$\lambda_1 = \lambda_\mathrm{band}+\Delta\lambda_\mathrm{band}+0.5\Delta\lambda$,
and we set $\Delta\lambda=0.1~\micron$.
Thus, $\delta_\mathrm{band}$ represents the difference in the average value of a transmission spectrum within a molecular absorption band relative to the average of the flanking regions.
We determine the value of $\delta_\mathrm{band}$ for each of the bands\footnote{The longest-wavelength bands of H$_{2}$O and O$_{3}$ are within $0.1~\micron$ of the long-wavelength end of the contamination spectra, so these two bands have truncated baselines for determining $\epsilon_\mathrm{cont}$.} 
illustrated in Figure~\ref{fig:contamination_spectra} and calculate the molecular offset $\delta_\mathrm{mol}$ for each molecule as the average of $\delta_\mathrm{band}$ for the molecular bands weighted by the band widths.

We find that none of the molecular offsets for CH$_{4}$, CO, CO$_{2}$, H$_{2}$O, N$_{2}$O, O$_{2}$, or O$_{3}$ are larger than our adopted 30~ppm detection threshold.
The largest offsets are those for O$_{2}$, H$_{2}$O, and CH$_{4}$, which are illustrated in Figure~\ref{fig:molecular_offsets}.
While all are still below the adopted detection threshold, the later spectral types produce relatively larger offsets.
For the \spots{} models, the O$_{2}$ and H$_{2}$O offsets are positive, while the CH$_{4}$ offsets are negative.
The offsets trend similarly in the \spotsfaculae{} models, except that the O$_{2}$ and H$_{2}$O offsets start to become more negative for spectral types later than around K5V, while those for CH$_{4}$ become more positive.
In each case, O$_{2}$ and H$_{2}$O offsets trend in the opposite direction as the CH$_{4}$ offsets.

Of course, these offsets are calculated for typically active FGK dwarfs and nominal transit depths of 1\%, so more active host stars or deeper transit depths could render the molecular offsets larger than our adopted detection threshold.
By the same token, improvements in observational techniques or instrumentation could enable finer precisions in transmission spectra than 30~ppm.
In any case, we point out here the trends in these molecular offsets so that they may be useful for identifying stellar contamination features in transmission spectra in future studies.

\subsubsection{Water spectral features} \label{sec:water}

Water features in transmission spectra are of particular interest, given their ubiquity in existing hot Jupiter \citep[e.g.,][]{Sing2016} and some hot super-Neptune \citep{Fraine2014, Stevenson2016, Wakeford2017} observations to date.
For typically active FGK dwarfs, considering the \spots{} models, we find the largest offsets at H$_{2}$O absorption bands for spectral type K7V.
In this case, unocculted spots inflate a 1\% transit depth by $\delta_\mathrm{H_{2}O} = 7^{+11}_{-4}$~ppm.
For \spotsfaculae{} models, by comparison, the largest offset, $\delta_\mathrm{H_{2}O} = 9^{+8}_{-3}$~ppm, is found for K4 dwarfs.
Generalized for any transit depth, these values correspond to maximal values of $\epsilon_\mathrm{H_{2}O} = 1.0007^{+0.0011}_{-0.0004}$ for \spots{} models and $\epsilon_\mathrm{H_{2}O} = 1.0009^{+0.0008}_{-0.0003}$ for \spotsfaculae{} models.

In both cases, the net effect of unocculted heterogeneities is to increase transit depths, potentially mimicking a planetary water absorption feature.
The scale of the effect, however, is far smaller than that of planetary features that have been probed in transmission spectra to date.
For comparison, the commonly studied 1.4~$\micron$ water absorption band has an amplitude of a few hundreds of ppm for hot Jupiters \citep[e.g.,][]{Sing2016} and the hot Neptunes in which it has been yet detected \citep{Fraine2014, Wakeford2017}.
Furthermore, the observed stellar contamination signal scales with the nominal transit depth (following Equation~\ref{eq:D_obs}), so for transits of planets smaller than hot Jupiters and Neptunes---in which the planetary atmospheric signals will be smaller than the existing detections---the stellar contamination signal will be correspondingly smaller as well.
Thus, we conclude that stellar contamination at wavelengths of interest for H$_{2}$O---or CH$_{4}$, CO, CO$_{2}$, N$_{2}$O, O$_{2}$, or O$_{3}$, for that matter---is not problematic for transmission spectroscopy studies involving typically active FGK host stars.
For a discussion of these species in transmission spectra of Earth-like planets, see Section~\ref{sec:earth_analog}.
As always, stellar activity is an important caveat: special care should be taken in studies involving host stars with larger variability amplitudes than the medians tabulated in Table~\ref{tab:variabilities} or other indicators of stellar activity.

Another important caveat comes from a potential limitation of our approach, which is that we fix spot and facula temperatures to set values.
As we are investigating an already large parameter space, out of necessity we do not allow for a range of active region temperatures for a given spectral type.
However, a range of active region temperatures are likely present on a given star.
On the Sun ($T_\mathrm{eff}=5800$~K), for example, sunspot umbrae generally have temperatures of 3900--4800~K and penumbrae 5400--5550~K \citep[][and references therein]{Solanki2003}.
In this study we adopt $T_\mathrm{spot} = 4030$~K for G2 dwarfs, roughly in line with these values.
Nonetheless, sunspots as cool as $T_\mathrm{spot} \simeq 3200$~K have been observed.
These are notable because water forms in sunspots cooler than about 3900~K and represents the dominant opacity source in unusually cool sunspots \citep{Wallace1995}.
Therefore, adopting a fixed spot temperature may lead us to underestimate $\delta_\mathrm{H_{2}O}$ for spectral types G8V and earlier, for which we set $T_\mathrm{spot} > 3900$~K.
Still, the values of $\delta_\mathrm{H_{2}O}$ that we determine for spectral types G9V--K9V are roughly two orders of magnitude below the amplitudes of planetary water absorption features that have been detected to date.
This fact suggests that our top-level conclusions are likely not affected by fixing active region temperatures to set values, though we caution that more detailed investigations are warranted for specific observational cases in which the host star is relatively active or the expected scale of the planetary feature is smaller than in the existing detections.

\subsection{Earth-Sun analog systems}
\label{sec:earth_analog}

\begin{figure}[!tbp]
\includegraphics[width=\linewidth]{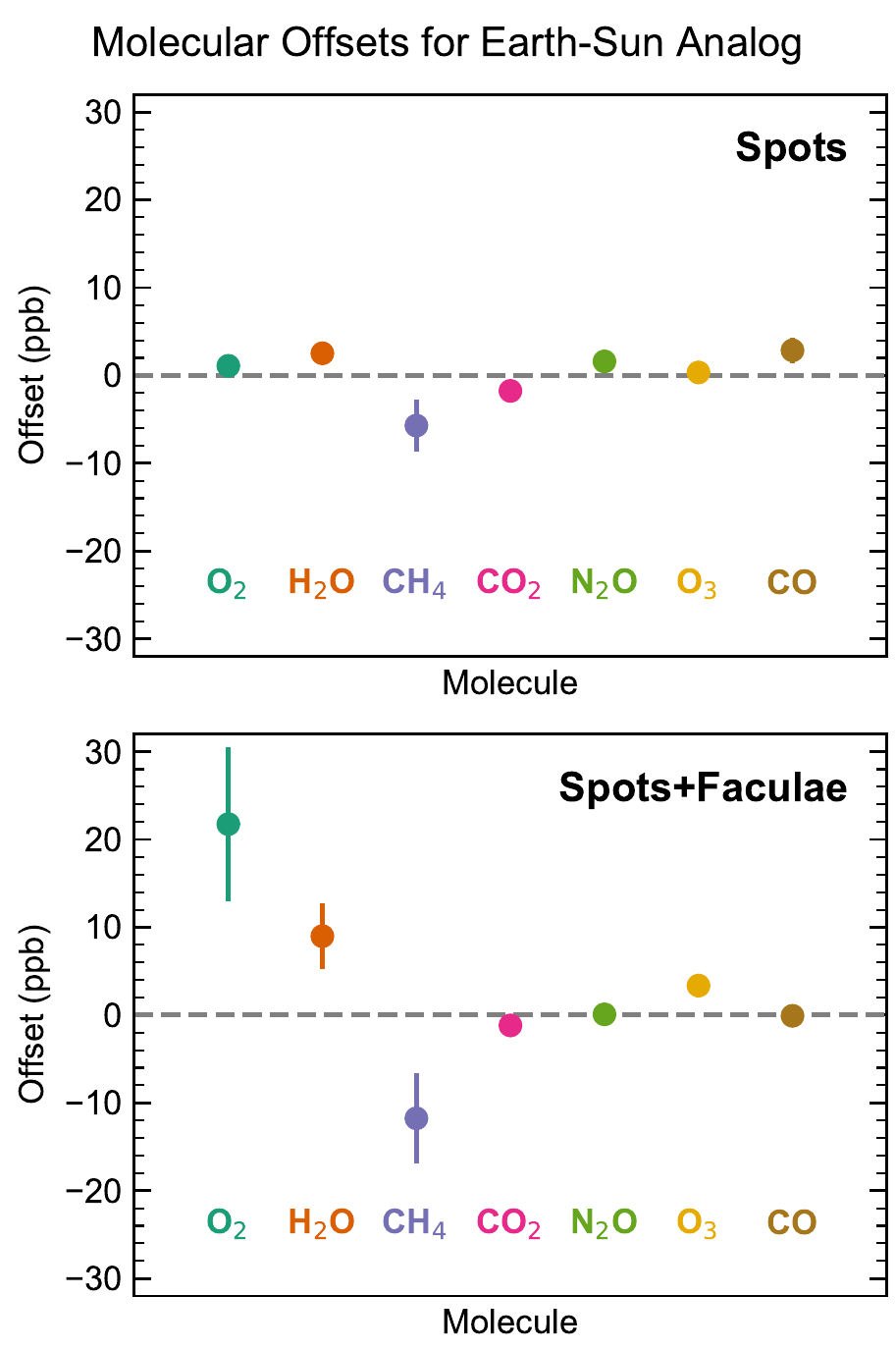}
\caption{Transit depth offsets within molecular absorption bands in transmission spectra of a typically active G2V dwarf, assuming a nominal 84~ppm transit depth. 
For both \spots{} (top) and \spotsfaculae{} (bottom) models, the offsets at wavelengths of interest for important planetary atmospheric species are more than an order of magnitude smaller than $\Delta D_{\earth}$.
The molecules are ordered by the the wavelength of their respective shortest-wavelength bands (see Figure~\ref{fig:contamination_spectra}), which illustrates that the largest offsets are generally found for the molecules with bands at shorter wavelengths. 
The error bars indicate $1\sigma$ prediction intervals, which are generally smaller than the point size.
Note that the offset values are given in parts-per-billion.
\label{fig:Earth-Sun}}
\end{figure}

One interesting example that warrants further investigation here is that of Earth-Sun analog systems.
These systems are targets of long-term efforts to characterize truly Earth-like exoplanets and search for biosignatures.
Given the Earth-Sun radius ratio, the nominal transit depth of such a system is $D_{\earth} = 84$~ppm.
Within the wavelength range of this study, Earth's transmission spectrum displays prominent absorption bands from from H$_{2}$O, CO$_{2}$, O$_{2}$, and O$_{3}$ \citep[e.g.,][]{Ehrenreich2006, Kaltenegger2009, Palle2009}.
An order of magnitude approximation for the scale of spectral features in transmission spectra $\Delta D$ \citep{Miller-Ricci2009} for an Earth-Sun system gives
\begin{equation}
\Delta D_{\earth} \sim \frac{2 H_{\earth} R_{\earth}}{R_{\odot}^{2}} = 2 \times 10^{-7}
\label{eq:deltaD}
\end{equation}
or 200 parts-per-billion (ppb) for features covering a single scale height.

For comparison, Figure~\ref{fig:Earth-Sun} illustrates the molecular offsets for important planetary molecular absorbers in the 0.05--5.5~$\micron$ range in an Earth-Sun analog system.
To calculate these offsets, we use the stellar contamination spectrum for the typically active G2 dwarf (presented in Figure~\ref{fig:contamination_spectra}) and assume $D=84$~ppm.
Of the molecular features highlighted in Figure~\ref{fig:contamination_spectra}, we find the largest overall offset, $\sim 20$~ppb, for O$_{2}$ with the \spotsfaculae{} model.
The remaining offsets are generally $<10$~ppb.
In other words, the scale of the stellar contamination is roughly an order of magnitude smaller than a single-scale-height planetary transmission feature.

We conclude, therefore, that stellar contamination in Earth-Sun analog systems will not preclude low-resolution observations of planetary molecular features.
High-resolution ($R\sim$~100,000) observations, in which planetary lines are Doppler-shifted away from stellar lines \citep[e.g.,][]{Snellen2010, Brogi2012, Rodler2012}, should suffer even less from this effect.
Given the future potential for high-resolution observations, including searches for potential biosignatures \citep{Snellen2013, Rodler2014, Ben-Ami2018}, a detailed examination of the effect of stellar contamination on high-resolution observations of Earth-Sun analog systems would be worthwhile, but it is outside of the scope of this work.
In any case, the minute scales of both $\Delta D_{\earth}$ and $\delta_{mol}$ emphasize the importance of precisely understanding the photospheric properties of interesting exoplanet host stars, including active region contrasts and covering fractions at the time of transit observations.

\subsection{TiO/VO in Visual Contamination Spectra}
\label{sec:TiO}

Titanium oxide (TiO) and vanadium oxide (VO) are two important molecular absorbers in planetary atmospheres, particularly in those of hot giant planets.
They display significant opacity across the full visual wavelength range \citep{Hill2013}, which allows them to significantly affect pressure-temperature profiles of hot giant planets.
Evidence for TiO/VO absorption features in the transmission spectra of the ultra-hot Jupiter \object{WASP-121b} \citep{Evans2016}, for example, pointed to the presence of a thermal inversion in the planetary atmosphere, which was later confirmed by an thermal emission spectrum obtained through secondary eclipse observations \citep{Evans2017}.

At the same time, TiO/VO are also present in stellar atmospheres.
They absorb more strongly at cooler stellar temperatures, and observations of TiO/VO molecular features have long been used to constrain spot temperatures and filling factors \citep{Vogt1979, Vogt1981, Ramsey1980}.
In this study, we find that unocculted spots can impart TiO/VO features in exoplanet transmission spectra.
This is most clearly illustrated by the K-dwarf contamination spectra in the lower left panel of Figure~\ref{fig:visual_spectra}, which closely resemble the absorption spectrum of TiO \citep{Hill2013}.

A straightforward calculation of $\delta_\mathrm{mol}$ for TiO and VO as defined in Section~\ref{sec:molecular_features} is complicated by the tight packing of molecular bands across the visual wavelength range, where their absorption cross-sections are important.
However, we can gain some quantitative insight into the impact of strong visual molecular absorbers in spots on transmission spectra by investigating deviations from simple slopes in the visual contamination spectra.
To this end, we define the TiO/VO offset as
\begin{equation}
\delta_\mathrm{TiO/VO} = \max{[ D (\epsilon - \epsilon_\mathrm{line}) ]},
\label{eq:delta_TiO}
\end{equation}
in which $\epsilon$ is the stellar contamination spectrum in the 0.4--0.9~$\micron$ range, $\epsilon_\mathrm{line}$ is a simple line fit to the points used in Section~\ref{sec:slopes} to define the visual offset (Equation~\ref{eq:delta_visual}), and we set $D=1\%$ as before.
In other words, $\delta_\mathrm{TiO/VO}$ provides an estimate of the amplitude of the deviations from a simple slope in a visual stellar contamination spectrum for a planet with a 1\% transit depth.
To simulate observational precisions, we calculate $\delta_\mathrm{TiO/VO}$ with stellar contamination spectra than have been down-sampled from the resolution of the PHOENIX models to a spectral resolution of $100$~\AA.

\begin{figure}[!htbp]
\includegraphics[width=\linewidth]{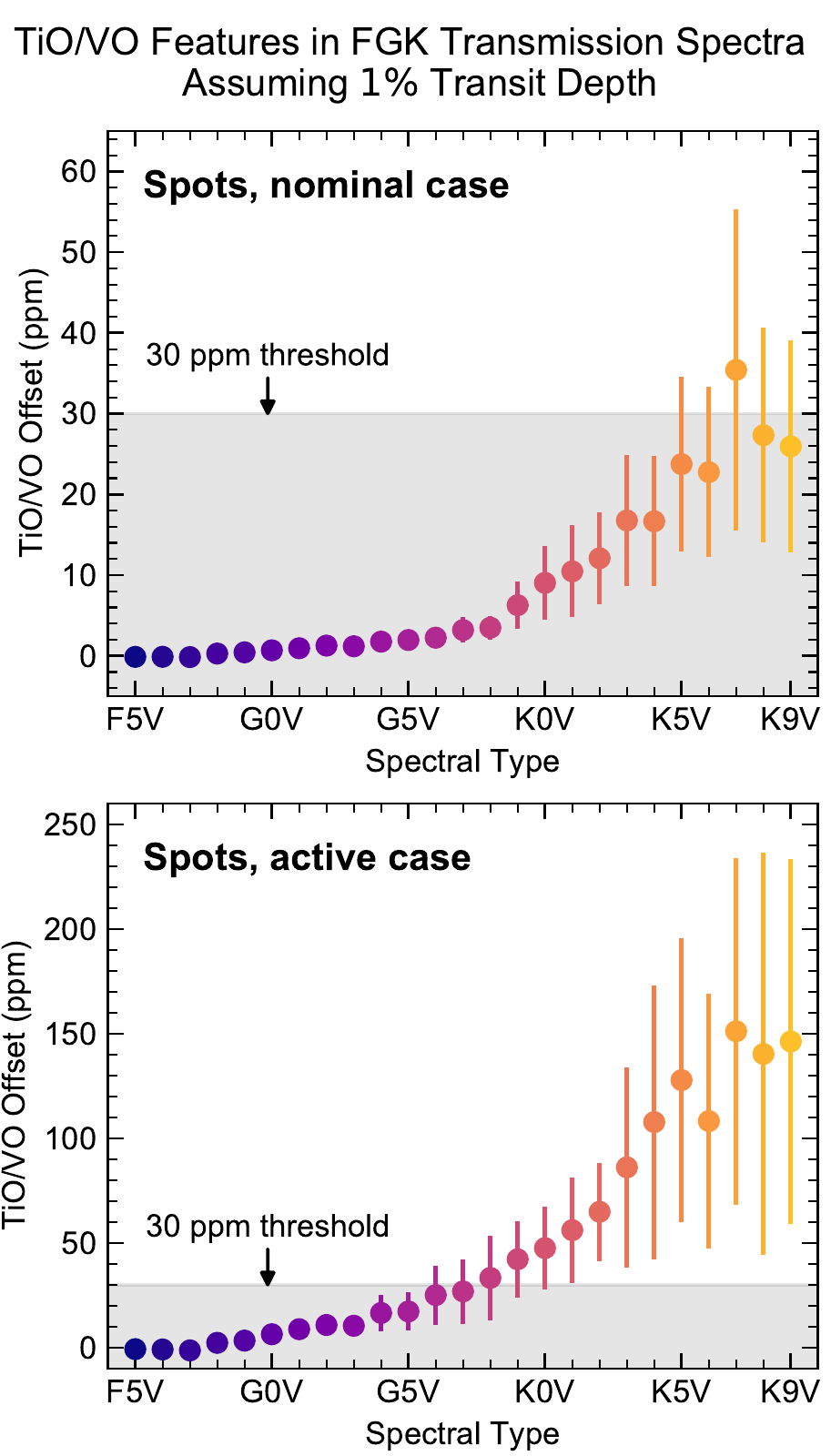}
\caption{TiO/VO offsets in transmission spectra for \spots{} models, assuming a nominal 1\% transit depth. 
The scale of the offsets grows with later spectral types.
For typically active FGK dwarfs (top), offsets are relatively small, reaching a few tens of ppm for late K dwarfs.
For more-active FGK dwarfs (bottom, see Section~\ref{sec:trends}), offsets are roughly 150~ppm for late K dwarfs.
The gray shaded region illustrates offsets that are below our adopted 30~ppm detection threshold.
\added{The data are color-coded by spectral type, following Figure~\ref{fig:fspot_variability}.}
Note the varying y-axis scales.
\label{fig:TiO_offsets}}
\end{figure}

We calculate $\delta_\mathrm{TiO/VO}$ for all \spots{} models, in which visual molecular features are most apparent, including contamination spectra and their $1\sigma$ prediction intervals for our nominals case and the active case defined in Section~\ref{sec:trends}.
The results are illustrated in Figure~\ref{fig:TiO_offsets}.
The offsets grow with later spectral types.
For our nominal case of typically active FGK dwarfs, we find that only K7 dwarfs produce offsets greater than our adopted 30~ppm detection threshold, though none of the estimates of $\delta_\mathrm{TiO/VO}$ are greater than a few tens of ppm.
For our active case, on the other hand, we find estimates for $\delta_\mathrm{TiO/VO}$ that are greater than 30~ppm for spectral types G8V and later and that are roughly 150~ppm for late K dwarfs.
We conclude that visual molecular features are generally not significant for typically active FGK dwarfs, though they can be significant for more-active K and late G dwarfs.
Therefore, we caution that stellar molecular features should be a consideration for late G and K dwarfs, especially if they display larger variability amplitudes than the medians tabulated in Table~\ref{tab:variabilities} or other indications of stellar activity.
Examples of such systems include \object{WASP-6}, \object{WASP-19}, and \object{HD~189733}, all of which are late G or early K dwarfs with relatively high chromospheric activity indices \citep[$\log{R^{'}_\mathrm{HK}} > -4.5$;][]{Sing2016}.

\subsection{Additional Impact of Stellar Chromospheres} \label{sec:chromospheres}

In this initial study we examine the effects of heterogeneity purely in a photospheric context.
However, we recognize the widespread occurrence of chromospheres in late-type stars, which may be operationally defined as an outer atmospheric region coinciding with the onset of a positive temperature gradient with height \citep{Linsky1980}.
In a physical context, chromospheric and coronal regions on the Sun and, by extension, in late-type stars are spatially associated with emergent magnetic flux, i.e., precisely the kind of heterogeneities that affect the interpretation of exoplanet transmission spectra.
Chromospheric heating can impact spectral line profile shapes and strengths, including those of key features such as the Ca~II H and K resonance lines in the blue-visible and their UV counterparts, the Mg~II h and k resonance lines; the Ca~II infrared triplet lines, the Na~I D lines and the Balmer lines.  
Additionally, lower chromospheric and upper photospheric heating can alter the ionization fractions of neutral metal species, notably that of the Fe I lines and K I in addition to the concentrations of molecular species such as CO.

Some quantitative insight on the magnitude of the effects of enhanced chromospheric heating on atomic lines is provided by results of long-term studies of solar variability.  
\citet{Livingston2007} summarizes observations of spectral line variability seen in the Sun-as-a-star in their multi-decadal program from 1974 to 2006.  
Inspection of the figures in \citet{Livingston2007} reveals, for example, that the peak-to-peak full disk cycle variations in the Ca~II K index, defined as the relative strength of the line core in a central 1 {\AA} bandpass, are approximately 25\%. 
The Na I D lines can change by 22\% in central intensity during the solar cycle, while photospheric Fe I lines can exhibit central intensity changes of $\sim$~6\%.  
The central depths of both the Ca~II IR triplet feature at 8542~{\AA} and H$\alpha$ also vary in phase with the solar cycle, though at lower relative amplitudes compared to the Ca~II K line.  

The particular case of the CO molecule is interesting because its formation and behavior is intimately linked to the inhomogeneous nature of the solar atmosphere.  
In particular, \citet{Ayres1981} observed that in the presence of localized mechanical heating, CO molecules begin to disassociate, leading to a decline in radiative cooling.  
A new equilibrium only is established at higher temperatures where ionized Ca and Mg become the dominant radiative coolants in the chromosphere. 
Outside of these regions, the outer atmospheres exists at a temperature less than that of the chromospheric temperature minimum with radiative cooling in the CO bands playing a key role in determining the local thermal structure.  
As \citet{Ayres1981} concludes, the heterogeneous solar atmosphere is thermally bifurcated between hot chromospheric regions where the CO molecule is depleted and locally cold regions where CO is enhanced \citep[see also][]{Ayres1996}.

At cooler effective temperatures beginning with the K dwarfs, H$\alpha$ becomes a prominent indicator of the presence of chromospheres in emission and absorption \citep{Cram1979, Cram1987}.  
The Ca~II core emission and H$\alpha$ strength are correlated with K dwarfs that exhibit very weak H and K emission and also show weak H$\alpha$ absorption that is dominated by the photospheric contribution.  
However, among late K dwarfs and early M dwarfs, even those objects with weak Ca~II emission still display significant H$\alpha$ absorption \citep{Robinson1990}. 
Thus, as \citet{Cram1987} conclude, the presence of H$\alpha$ chromospheres in K and M dwarfs is ubiquitous---a truly immaculate star in this class may be nonexistent.  
Therefore, future investigations of specific atomic or molecular features as they may appear in exoplanet transmission spectra may have to include considerations of the impact of chromospheric and coronal heating on their formation, depending on the level of precision required.

\subsection{Promising Paths Forward}
\label{sec:paths_forward}

While we find that stellar contamination in transmission spectra of FGK dwarfs is less problematic generally than found for M dwarfs in \ponet{}, there are still circumstances when observers should tread carefully.
In particular, special care should be taken to disentangle stellar and planetary features in observations involving mid-G to late-K dwarfs---especially active ones---and minute planetary spectral features on the order of tens of ppm or less.
Here we briefly review approaches that can be useful in these situations.

There are a suite of forward-modeling approaches that provide useful priors for interpreting transmission spectra.
In particular, we use variability models in this work to explore spot and facula covering fractions for typically active FGK dwarfs and their associated range of stellar contamination signals.
These results can be applied to appropriate FGK host stars, i.e. those with variability amplitudes comparable to the medians tabulated in Table~\ref{tab:variabilities}.
For more or less active stars, the scaling relation coefficients provided in Table~\ref{tab:scaling_coeffs} can be used to estimate the spot covering fraction, which in turn can be used to approximate the scale of the stellar contamination signal relative to those detailed here.
For simplicity, we present observational offsets in Section~\ref{sec:discussion} assuming $D=1\%$, but these values all scale directly with $D$, so it is trivial to scale them to different transit depths.

The same general forward-modeling approach can be applied to individual interesting stars.
\citet{Spake2018}, for example, apply the approach detailed here to \object{WASP-107} and find that the scale of the observed helium absorption feature at 10,833~{\AA} in the transmission spectrum of \object{WASP-107b} is much greater that what can be produced by photospheric heterogeneities.
These authors also investigate and discount the possibility that the observed helium feature could arise from an inhomogeneous chromosphere, which is an important step for attributing a planetary origin to lines that are also present in chromospheres \citep[see also][]{Cauley2018}.

When applying this approach to individual host stars, active region crossings observed during exoplanetary transits are particularly helpful.
These light curve anomalies encode the active region size and contrast (i.e., temperature), estimates of which can be obtained with tools like \texttt{SPOTROD} \citep{SPOTROD} or \texttt{PyTranSpot} \citep{PyTranSpot}.
These parameters in turn provide useful inputs to the variability modeling approach that we employ here, refining estimates of the total active region covering fractions corresponding to an observed photometric variability \citep[e.g.,][]{Espinoza2019}.

Even more detailed studies of important individual stars can provide further insights.
For example, using a combination of high-resolution NIR spectra and long-term photometric monitoring, \citep{Gully-Santiago2017} constrain the spot temperature of the weak-lined \textit{T-Tauri} star \object{LkCa 4} and trace the temporal evolution of the spot filling factor.
Combining both radial velocity and photometric time-series, the \texttt{StarSim} tool \citep{StarSim} can also be used to trace the temporal evolution of photospheric heterogeneities and thus the stellar contamination signals at the  time of transit.
Studies of the out-of-transit stellar spectra flanking transit observations can provide further insights into the relative change in the stellar contamination signal between transits \citep{Zellem2017}.

Finally, transmission spectra retrievals that allow for stellar contamination can be used to disentangle stellar and planetary spectral features.
Within a nested sampling framework \citep{Skilling2006}, the Bayesian evidence for models with and without stellar contamination can be straightforwardly compared.
Using this approach, \citet{Espinoza2019} concluded that the TiO absorption features observed in the visual transmission spectrum of \object{WASP-19b} are likely produced by unocculted spots in the photosphere of the active G9V host star.
Meanwhile, using the same approach, Bixel et al. (submitted) found no evidence of stellar contamination in the visual transmission spectrum of \object{WASP-4b}, a system similar in most respects but with a less-active host star.
In a systematic study of the \citet{Sing2016} sample of hot Jupiters using a joint stellar and planetary retrieval framework, \citet{Pinhas2018} identified a tentative but suggestive trend between the chromospheric activity index $\log{R^{'}_\mathrm{HK}}$ and the Bayesian evidence in support of models that allow for stellar contamination features.
If confirmed, this finding suggests that $\log{R^{'}_\mathrm{HK}}$ can be used to predict whether stellar contamination will affect transmission spectra from a given host star.
Along with the trends in stellar contamination features discussed in Section~\ref{sec:trends}, systematic trends like these can provide further context for interpreting spectral features in a given transmission spectrum.

\section{Conclusions} \label{sec:conclusions}

We have presented a study of photospheric heterogeneity in FGK stars and its associated effect on exoplanet transmission spectra in the 0.05--5.5~$\micron$ wavelength range.
The key results of this study are as follows:

\begin{enumerate}

\item 
For both \spots{} and \spotsfaculae{} models, rotational variability amplitudes in the \textit{Kepler} bandpass show a square-root-like dependence on the spot covering fraction, allowing estimates of spot covering fractions to be obtained from observed variabilities.

\item
Relative to M dwarfs, the lower variabilities that are typically observed for FGK stars point to lower active region covering fractions and enable tighter estimates on the covering fractions from rotational variability modeling.

\item 
We find that the median \textit{Kepler} variability amplitudes for spectral types F5V--K9V correspond to spot covering fractions that generally increase with later spectral types, from roughly 0.1\% for F dwarfs to 2--4\% for late K dwarfs.

\item
If present on the unocculted stellar disk, these heterogeneities 
\replaced{on transmission spectra primarily increase}
{primarily impact transmission spectra by increasing}
transit depths across the studied wavelength range.
The largest differences between the stellar contamination spectra that we calculate for \spots{} and \spotsfaculae{} models occur at wavelengths $\lessapprox 0.5~\micron$, for which the \spots{} models predict relatively large increases in transit depth, while the \spotsfaculae{} models predict strong decreases in transit depth.
Thus, transit observations at short wavelengths can be used to constrain the presence of unocculted faculae on the stellar disk.

\item
In general, the largest impacts of stellar contamination in transmission spectra are evident at UV and visual wavelengths.
We calculate the offsets between blue ($0.4~\micron$) and red ($0.9~\micron$) visual transit depths owing to stellar contamination.
Assuming a nominal transit depth of 1\% and a 30~ppm detection threshold, we find that typically active G and K dwarfs can impart detectable visual offsets on transmission spectra.

\item
Exploring line offsets in stellar contamination spectra around H$\alpha$ and the Na D and K doublets, we find that unocculted spots on typically active FGK dwarfs do not alter transit depths detectably, though unocculted faculae in K dwarfs can decrease transit depths around the Na D doublet by a few hundreds of ppm.
For more active host stars, we caution that detectable changes may be evident for more atomic features and earlier spectral types, and we suggest that trends in relative strengths of these features can be used to identify their stellar origin.

\item
We calculate transit depth offsets at wavelengths of interest for CH$_{4}$, CO, CO$_{2}$, H$_{2}$O, N$_{2}$O, O$_{2}$, and O$_{3}$ and find that none are detectable for typically active FGK dwarfs, again assuming a 1\% transit depth and 30~ppm detection threshold.
Of these, the largest offsets are apparent 
\added{at wavelengths of interest for}
O$_{2}$, H$_{2}$O, and CH$_{4}$, which have molecular bands at shorter wavelengths.
Larger offsets are possible for more active host stars, and so we suggest that future works exercise care when studying these features in the atmospheres of exoplanets hosted by active G and K stars.

\item
Defining the deviation of the visual stellar contamination spectrum from a simple slope as a proxy for TiO/VO features, we find that stellar TiO/VO features in transmission spectra are potentially detectable for typically active late-K dwarfs and, for active stars, can be apparent for spectral types as early as G8V.

\item
Taking the long view, we explore stellar contamination in an Earth-Sun analog system and find that transit depth offsets due to stellar contamination at wavelengths of interest for important atmospheric molecular absorbers are $\lessapprox 20$~ppb, roughly an order of magnitude less than the scale of a planetary atmospheric feature covering a single scale height.

\end{enumerate}

The whole of this analysis shows that stellar contamination in transmission spectra of FGK-hosted exoplanets is generally less problematic than for exoplanets orbiting M~dwarfs.
\added{The impact of the TLS effect is most prominent at shorter wavelengths.
While it can produce detectable slopes in visual transmission spectra from G and K dwarfs and, for more-active late-G and K dwarfs, detectable offsets at wavelengths of interest for TiO/VO, TLS signals are generally minor at wavelengths of planetary atomic and molecular features.}
This bodes well for high-precision observations of these targets, including those expected to be discovered by the recently launched \textit{TESS} mission \citep{Ricker2015}, with 
\added{current ground- and space-based facilities and}
near-future facilities like 
\replaced{the \textit{James Webb Space Telescope}}
{\textit{JWST}}.
However, within the parameter space that we explore, more care should be exercised for observations at shorter wavelengths and those with host stars that are more active or of later spectral types.

\acknowledgments

B.R. acknowledges support from the National Science Foundation Graduate Research Fellowship Program under Grant No. DGE-1143953. 
D.A. acknowledges support from the Max Planck Institute for Astronomy, Heidelberg, for a sabbatical visit. 
M.S.G. thanks the Lunar and Planetary Lab at the University of Arizona for hosting him during his sabbatical leave.
\added{We thank the anonymous referee for their constructive comments.}
The results reported herein benefited from collaborations and/or information exchange within NASA's Nexus for Exoplanet System Science (NExSS) research coordination network sponsored by NASA's Science Mission Directorate. 
The National Solar Observatory is operated by AURA under a cooperative agreement with the National Science Foundation.
This research has made use of NASA's Astrophysics Data System.

%



\software{Astropy \citep{Astropy2013, Astropy2018}, 
Matplotlib \citep{Matplotlib}, 
NumPy \citep{NumPy}, 
SciPy \citep{SciPy}}




\bibliography{references}



\end{document}